\documentclass[twocolumn, showpacs, amsmath, amssymb, prb]{revtex4}

\usepackage{amssymb}
\usepackage{amsmath}
\usepackage{amsfonts}
\usepackage{graphicx}

\makeatother

\newcommand{\be}{\begin{equation}}
\newcommand{\ee}{\end{equation}}
\newcommand{\ba}{\begin{eqnarray}}
\newcommand{\ea}{\end{eqnarray}}

\begin{document}

\title{Optical and transport properties of spheroidal metal nanoparticles with account for the surface effect}

\author{Nicolas I.~Grigorchuk \footnote{email: ngrigor@bitp.kiev.ua}}

\affiliation{Bogolyubov Institute for
Theoretical Physics, National Academy of Sciences of Ukraine, \\
 14-b Metrologichna Str., Kyiv-143, Ukraine, 03680}

\author{Petro M. Tomchuk \footnote{email: ptomchuk@iop.kiev.ua}}

\affiliation{Institute for Physics, National Academy of Sciences of Ukraine, \\
 46, Nauky Ave., Kyiv-28, Ukraine, 03680}

\begin{abstract}
The kinetic approach is applied to develop the Drude-Sommerfeld
model for studying of the optical and electrical transport properties of
spheroidal metallic nanoparticles, when the free electron path is much greater
than the particle size. For the nanoparticles of an oblate or a prolate
spheroidal shape there has been found the dependence of the dielectric function
and the electric conductivity on a number of factors, including the frequency,
the particle radius, the spheroidal aspect ratio and the orientation of the electric field
with respect to the particle axes. The oscillations of the real and imaginary parts of the
dielectric permeability have been found with increasing of particle size at some fixed
frequencies or with frequency increasing at some fixed radius of a nanoparticle.
The results obtained in kinetic approach are compared with the known from the classical model.
\end{abstract}


\date{\today}

\maketitle


\newpage
\section{INTRODUCTION}

 Understanding how light interacts with matter at the nanometre scale is a fundamental problem in nanophotonics and optoelectronics.
 The variety of applications are based on the surface-bound optical excitations in metallic nanoparticles (MNs). They include plasmonic nanomaterials, optical nanosensors, biomarkers, integrated circuits and sub-wavelength waveguides, molecular imaging, metamaterials
 with a negative refractive index, optical antennas and so on (see, e.g., Refs.[\onlinecite{ZSS}--\onlinecite{RM}]).

 The optical and transport properties of the MNs have been investigated for a long time and the results have been presented
 quite completely in numbers of reviews\cite{HarR,Hee,Hal,PerR} and monographs,\cite{Kl,NC,BH,KV,SB}
 especially for MNs of a spherical shape.

 Dielectric function is the most important factor for the design and optomization of plasmon nanometer-sized structures.
 However, this function for MNs differs from an ideal bulk metal. The difference depends on many factors, such as the size and shape
 of the MNs and the surrounding media.\cite{LMS,KCZ,KP,TG,CeC} That is why the dielectric function of MNs has been under comprehensive study for many years. \cite{Kl,NC,BH,KV,SB,KK} The size effect becomes more significant when the size $d$ of MN is comparable with the electron mean free path $l$. The ratio between $d$ and $l$ is a very important physical characteristic, which eventually defines the mechanism of an electron relaxation rate inside the MN. As a rule, the cases $d\gg l$ and/or $d\ll l$ are studied. The former case refers to the so called diffusive electron dynamics and the latter one to the ballistic electron dynamics. The diffusive case has been studied in details, since it enables to use the Mie theory for a uniform media\cite{Mie} or the Maxwell-Garnett theory\cite{Max} for a composite media in the calculations of the optical properties of the MN.

 In particular, when $d\gg l$, the following expression for the dielectric permeability which
 results from the Drude-Sommerfeld theory\cite{Dru,SB} has been often used:
\begin{equation}
 \label{eq 1}
  \epsilon(\omega)=\epsilon'(\omega)+ i \epsilon''(\omega)=1-\frac{\omega^2_{pl}}
   {\nu^2+\omega^2}+i\frac{\nu}{\omega}\frac{\omega^2_{pl}}{\nu^2+\omega^2}.
    \end{equation}
Here $\nu$ is the collision frequency inside the particle bulk, $\omega_{pl}=\sqrt{4\pi n e^2/m}$ is the frequency of plasma
electrons oscillations in the metal, $e$ and $m$ are the electron charge and mass, correspondingly, and $n$ is the electron concentration. The imaginary part of the dielectric permeability $\epsilon''(\omega)$ is connected with the high-frequency (optical) conductivity by the well known relation:
\begin{equation}
 \label{eq 2}
  \sigma(\omega)=\frac{\omega}{4\pi}\epsilon''(\omega)=\frac{\nu}{4\pi}\frac{\omega^2_{pl}}{\nu^2+\omega^2}.
   \end{equation}
 For sufficiently low frequencies ($\omega\rightarrow 0$), one gets the expression $\sigma_0=ne^2/m\nu$ describing the statical conductivity.

 In the case when the sizes of the MN are less than the mean electronic free path in the particle, the mechanism of electrons scattering
 is changed, and the surface of the particle starts to play a dominate role. This effect will be called further as the surface or boundary effect. Strictly speaking, in this case neither Mie nor Drude-Sommerfeld theory can be applied and an another approach ought
 to be elaborated to present the optical properties of MNs. As it was shown,\cite{Dor,SND} a clear size dependence of relaxation dynamics is observed in experiments. This is a strong indication of an efficient electron-surface phonon interaction in this regime. But the experimental results\cite{LMS,AA,Ber,GHN,HFM,HFW,RGM} are often analyzed within the frame of the named theories, or to simplify the problem, the Eqs.~(\ref{eq 1}), (\ref{eq 2}) have been used with formal replacements\cite{KV,HFW,KS,AKS}
\begin{equation}
 \label{eq 3}
  \nu\rightarrow\frac{3}{4}\frac{\upsilon_F}{R},\qquad \textrm{or} \qquad\nu\rightarrow A\upsilon_F\frac{S}{4V},
   \end{equation}
where $\upsilon_F$ is the electron velocity at the Fermi surface, $R$ refers the particle radius, $V$ and $S$ refer the volume and the surface area of the spherical particle, and $A$ is a coefficient obtained by fitting the calculations to the experimental data. In other words, the collision frequency in the particle volume is replaced by some effective collision frequency. But such a replacement can be applied for the MNs of a spherical shape only. If the geometry of MN differs from the spherical one, and the condition $d\ll l$ holds for at least one of the particle directions, then the optical conductivity becomes the tensor quantity \cite{TG} and the formal replacement similar to the one presented by Eq.~(\ref{eq 3}) can not
be used any more. It is necessary to look more closely at the effect of the particle boundaries on the optical properties. For a theoretical study it is convenient to choose the particle of a spheroidal shape. That is because the results obtained for the particles of such a shape can be easily extended to the particles of other shapes by means of the formal transformation of the spheroidal axes.

Thus, there arises a problem in the case $d\ll l$, how to calculate the values of $\epsilon$ or $\sigma$, when the mentioned above classical models can not be used.

The present work is devoted to the elaboration of the new approach, allowing to calculate the real and imaginary parts of the dielectric permeability for MNs of a spheroidal shape in the case when the inequality $d\ll l$ is hold.

Since both the real and the imaginary parts of $\epsilon(\omega)$ govern the numerous properties of the MNs,
the necessity of the detailed study of the boundary effect on the dielectric permeability has become evident.

The rest of the paper is organized as follows. The kinetic approach to the problem is presented in Sec. II. Section III contains the study of the conductivity in the spheroidal MNs. In Sec. IV, we consider the limit cases of the problem and the size effects. Sec. V is devoted to the discussion of the obtained results, and Sec. VI contains the conclusions.

\section{KINETIC EQUATION METHOD}

To account for the effect of MN boundaries on the optical and electrical transport properties of the MN, we will apply the kinetic equations approach. The advantages of this approach is that the obtained results can be applied to strongly anisotropic spheroidal (needle-like or disk-shaped) MNs, but in the case of MNs of spherical shape it transforms to the well known results\cite{Dru,SB},
like the ones given by Eqs.~(\ref{eq 1}), (\ref{eq 2}). Thus, it permits one to study the effect of the particle shape on the measured physical values. Secondly, the kinetic method enables to investigate the MNs with sizes as more or less then the electron mean-free
path $l$. But, there exists the lower limit of the applicability of this method in the small radius limit, when the particle size is comparable with the de Broglie wavelength of the electron, and the quantization of the electron spectrum start to play an essential role.\cite{YB,MWJ,WMW} Practically, it is around a radius of more or less 2 nm. Beyond the electronic Fermi distribution function, the influence of quantum effects on the optical conductivity can manifest itself in the quantization of electron pulses and angular momentums. It is important at low temperatures, when the distance between successive energy levels is much higher than $k_B T$.

Let us consider the single MN which is irradiated by electromagnetic wave, whose electric field is given as
\begin{equation}
 \label{eq 4}
  {\bf E}={\bf E}_0 \exp[i({\bf kr}-\omega t)].
   \end{equation}
Here ${\bf E}_0$ is the amplitude of an electric field, $\omega$ is its frequency,
${\bf k}$ is the wave vector, and ${\bf r}$ and $t$ describe the spatial coordinates and time.

We will assume that the electromagnetic wave length is far above the particle size. If one chooses
the coordinate origin in the center of the particle, then the above mentioned assumption is written as
\begin{equation}
 \label{eq 5}
  kr\ll 1.
   \end{equation}

The inequality (\ref{eq 5}) implies that the ${\bf E}$-field of the electromagnetic wave can be considered as spatially uniform on scales of the order of a particle size. This means that the field represented by Eq.~(\ref{eq 4}) induces inside of the MN the electric field varying in time, but uniform in space. The amplitude of such a field is connected with ${\bf E}_0$ by the relation\cite{GT}
\begin{equation}
 \label{eq 6}
  E_{in}^j(\omega)=\frac{E_0^j(0,\omega)}{1+L_j [\epsilon(\omega)-1]},
   \end{equation}
where $L_j$ are depolarization factors in the $j$-th direction (in the principal axes of an ellipsoid). The explicit expressions
 of $L_j$ for a single MN of an ellipsoidal shape can be found elsewhere (see, e.g., Refs. [\onlinecite{Osb},\onlinecite{LL}]).

The field $E_{in}$ has an effect on the equilibrium electron velocities distribution and, thus, determines the
appearance of a nonequilibrium addition $f_1({\bf r, v},t)$ to the Fermi distribution function $f_0(\varepsilon)$.
Here $\varepsilon=m\upsilon^2/2 $ is the kinetic energy of electron, and $\upsilon=|{\bf v}|$ refers the electron velocity.
As is well known,\cite{Ruch} the equilibrium function $f_0(\varepsilon)$ does not give any input to the current. With the account
for both the time dependence of Eq.~(\ref{eq 4}) and the inequality (\ref{eq 5}), the distribution function of electrons, which
generates the field $E_{in}$, can be written as

\begin{equation}
 \label{eq 7}
  f({\bf r,v},t)=f_0(\varepsilon)+f_1({\bf r,v},t)\equiv f_0(\varepsilon)+f_1({\bf r,v})\,e^{i\omega t}.
   \end{equation}
The function $f_1({\bf r, v})$ can be found as a solution of the linearized Boltzmann's equation

\begin{equation}
 \label{eq 8}
  (\nu-i\omega) f_1({\bf r, v})+{\bf v}\frac{\partial f_1}{\partial
   {\bf r}}+e{\bf E}_{in}{\bf v}\frac{\partial f_0(\varepsilon)}{\partial\varepsilon} =0.
    \end{equation}
In Eq.~(\ref{eq 8}) we have assumed that the collision integral
$(\partial f_1/\partial t)_{col}=-f_1/\tau$ is evaluated in the relaxation time approximation ($\tau=1/\nu$).
What is more, the function $f_1({\bf r,v})$ ought to satisfy the boundary conditions as well.
These conditions may be chosen from the character of electrons reflection from the inner walls of the MN.
We will take, as is usually done, the assumption of a diffusive electron scattering by the boundary of MN.
Placing the origin of the coordinates at the center of the particle, we can present this boundary conditions in the form
\begin{equation}
 \label{eq 9}
  f({\bf r,v})|_S =0, \qquad {\bf v}_n<0,
   \end{equation}
where ${\bf v}_n$ is the velocity of the component normal to the particle surface.

Alongside with diffusive, the mirror boundary conditions at the nanoparticle surface were examined in literature for electron scattering (see, e.g., Refs. [\onlinecite{Ruch}, \onlinecite{BKY}]). In this case, each electron is reflected from the surface at the same angle at which it falls to the surface. In diffuse reflection, the electron is reflected from the surface at any angle. In order that the mirror mechanism be dominant, the surface must be perfectly smooth in the atomic scale, since the degree of reflectivity of the boundary essentially depends on its smoothness. Practically, for a nonplanar border such smoothness is extremely difficult to achieve. As it was shown\cite{BKY}, the mirror boundary conditions give a small corrections to the results obtained with the account of only the diffusive electron reflections. Therefore, we chose more realistic boundary conditions given by Eq.~(\ref{eq 9}).

It is comparatively easy to solve Eq.~(\ref{eq 8}) and to satisfy the boundary conditions of Eq.~(\ref{eq 9}), if one passes to the transformed coordinate system, where an ellipsoid with semiaxes $R_1, R_2, R_3$ transforms into a sphere of radius $R$ with the same total volume:
\begin{equation}
 \label{eq 10}
  x_j =\frac{R_j}{R}x'_j, \qquad R=(R_1 R_2 R_3)^{1/3}.
   \end{equation}
Similar transformation should be made for the electron velocities as well: $\upsilon_j=\upsilon'_j R_j/R$.
Then solving Eq.~(\ref{eq 8}), one finds
\begin{equation}
 \label{eq 11}
  f_1({\bf r,v,}\,t)=-e\frac{\partial f_0}{\partial\varepsilon}
   {\bf v E}_{in}\frac{1-\exp[-(\nu-i\omega)\;t'({\bf r',v'})]}{\nu-i\omega},
    \end{equation}
where the characteristic of Eq.~(\ref{eq 8}) can be presented as

\begin{equation}
 \label{eq 12}
  t'=\frac{1}{{\bf v}^{'2}}\left[\bf r'v'+\sqrt{(R^2-r^{'2}) \,{\bf v}^{'2}+({\bf r' v'})^2}\right].
   \end{equation}
The characteristic curve of Eq.~(\ref{eq 12}) depends only on the absolute value of ${\bf R}$ and does not depend on the direction of ${\bf R}$. The radius vector ${\bf R}$ determines the starting position of an electron at the moment $t'=0$.

Generally speaking, the presence of the surface changes both the current and the field distributions.
It is reasonable to point out here that, though the electric field still remains homogeneous inside of the MN (in accordance
with the Eq.~(\ref{eq 5})), the distribution function $f_1$ is the coordinate dependent in any case due to the necessity
to meet the boundary condition given by Eq.~(\ref{eq 9}).

Performing the Fourier transformation of Eq.~(\ref{eq 11}), one can calculate the density of a high-frequency
current induced by the electromagnetic wave of Eq.~(\ref{eq 4}) inside the MN, via the expression
\begin{equation}
 \label{eq 13}
  {\bf j}({\bf r},\omega)=2e\left(\frac{m}{2\pi\hbar}\right)^3\int\int\int {\bf v} f_1({\bf r,v,}\,\omega)\,d^3 \upsilon.
   \end{equation}

Let us introduce the tensor of the complex conductivity $\sigma^c_{\alpha\beta}({\bf r}, \omega)$
using the relationship
\begin{equation}
 \label{eq 14}
  j_{\alpha}({\bf r},\omega)=\sum\limits_{\beta=1}^{3}\sigma_{\alpha\beta}^c({\bf r,\omega})\,E_{in}^{\beta}.
   \end{equation}
Then in accordance with the both Eq.~(\ref{eq 11}) and  Eq.~(\ref{eq 13}),
the components of this tensor can be presented in the form
\begin{eqnarray}
 \label{eq 15}
  \sigma_{\alpha\beta}^c({\bf r,\omega})&=&2e\left(\frac{m}{2\pi\hbar}\right)^3\int\int\int
   \upsilon_{\alpha}\left[-e\upsilon_{\beta}\frac{\partial f_0}{\partial\varepsilon}\right.
    \nonumber \\ &\times &
     \left.\left(\frac{1-e^{-(\nu-i\omega)t'(r',\upsilon')}}{\nu-i\omega}\right)\right]\;d^3\upsilon.
      \end{eqnarray}

  Before the detailed study of the role of the particle surface and the size effects, which will be displayed below, it is worth to note here the following. The surface effect on the conducting phenomenon is described in Eq.~(\ref{eq 15}) by means of the characteristic $t'({\bf r', v'})$. It accounts for the restrictions imposed on the electron movement by a nanoparticle surfaces. As one can see from Eq.~(\ref{eq 12}), the value of $t'$ is of the order of $t'\sim R/\upsilon_F$, where $\upsilon_F$ is the Fermi velocity. This implies that the value reciprocal to $t'$ will correspond to the vibration frequency between the particle walls. Hence, the inequality $\nu t'\gg 1$ indicates that the electron collision frequency inside the volume of MN would significantly exceed the one for an electron collision with the surface of MN. If pointed inequality is satisfied, it is possible to direct $t'\rightarrow\infty$ and the exponent in Eq.~(\ref{eq 15}) can be neglected. Then, we obtain a standard expression for the dielectric permeability, such as given in Eq.~(\ref{eq 2}). To ensure in that, it is necessary to pass in Eq.~(\ref{eq 15}) to the integration over $\upsilon$ in the spherical coordinate system
\[
\int\!\!\!\int\limits_{-\infty}^{\infty}\!\!\!\int\;d^3\;\upsilon\rightarrow\int\limits_{0}^{2\pi}d\varphi
\int\limits_0^{\pi}\sin\theta\,d\theta\int\limits_0^{\infty}\upsilon^2\,d\upsilon,
\]
with the use of the following formulas
\begin{equation}
 \label{eq 16}
  \int_0^{\infty}\upsilon^4\,\delta(\upsilon^2-\upsilon^2_F)\, d\upsilon=\frac{\upsilon^3_F}{2},
   \end{equation}
\begin{equation}
 \label{eq 17}
  n=\frac{8\pi}{3}\left(\frac{m\upsilon_F}{2\pi\hbar}\right)^3,
   \end{equation}
and take into account that the energy derivative of $f_0$ in zero approximation
in the small ratio of $k_B T/\varepsilon_F$, can be replaced by
\begin{equation}
 \label{eq 18}
  \frac{\partial f_0}{\partial\varepsilon}\approx -\delta(\varepsilon-\varepsilon_F),
   \end{equation}
as well as the fact that only the diagonal terms with $\upsilon_{\alpha}=\upsilon_{\beta}=\upsilon$
are retained after integration over all angles.

At the end of this section, we would like to pay attention on the next two important circumstances. i) Though the inner field ${\bf E}_{in}$ is spatially uniform, the distribution function $f_1({\bf r,v,t})$, and, consequently, the density of the current ${\bf j}({\bf r},\omega)$ are depended on the coordinates. This dependence is imposed by the boundary conditions of Eq.~(\ref{eq 8}). ii) The density
of the current and the components of the conductivity tensor have the physical sense only if they are averaged over the particle volume.
For instance, it is easy to ensure that the energy absorbed by a single MN is determined either by an averaged density of the current $\langle j\rangle$ or by an averaged tensor of the complex conductivity $\langle \sigma_{\alpha\beta}^c({\bf r},\omega)\rangle$
(with the account of the Eq.~(\ref{eq 14})).

\section{CONDUCTIVITY OF SPHEROIDAL METALLIC NANOPARTICLE}

Let us average over coordinates the components of the conductivity complex tensor represented by Eq.~(\ref{eq 15}).
The necessity of such an averaging arises from the fact that the power absorbed by a single MN is caused by the conductivity
averaged over the volume of MN. Then, in view of Eq.~(\ref{eq 16}), one gets the expression
\begin{eqnarray}
 \label{eq 19}
  &\langle\sigma_{\alpha\beta}^c({\bf r,\omega})\rangle = \frac{4 e^2}{m} \left(\frac{m}{2\pi\hbar}\right)^3
   \frac{1}{\nu-i\omega}\int\frac{d^3 r}{V}& \nonumber \\&\times
    \int d^3\upsilon\,\upsilon_{\alpha}\upsilon_{\beta}\,\delta(\upsilon^2-\upsilon^2_F)\left[1-e^{(\nu-i\omega)t'}\right]&,
     \end{eqnarray}
where the angle brackets denotes the averaging.
Firstly, we can fulfil the integration in Eq.~(\ref{eq 19}) over all electron coordinates.
In accordance with Ref. [\onlinecite{GT}], one can find that
\begin{equation}
 \label{eq 20}
  \frac{1}{V}\int\,d^3 r'\left[1-e^{-(\nu-i\omega)t'({\bf r',\upsilon'})}\right] = \frac{3}{4}\Psi(q),
   \end{equation}
where the following notations have been used
\begin{equation}
 \label{eq 21}
  \Psi(q)=\frac{4}{3}-\frac{2}{q}+\frac{4}{q^3}-\frac{4}{q^2}\left(1+\frac{1}{q}\right)e^{-q},
   \end{equation}
\begin{equation}
 \label{eq 22}
  q=\frac{2R}{\upsilon'}(\nu-i\omega)\equiv q_1-iq_2.
   \end{equation}
Further, we will take into account only the diagonal components of the conductivity tensor.
Based on Eq.~(\ref{eq 20}), we can rewrite Eq.~(\ref{eq 19}) as
\begin{eqnarray}
 \label{eq 23}
  \langle\sigma_{\alpha\alpha}^c({\bf r,\omega})\rangle & = &\frac{3e^2 m^2}{(2\pi\hbar)^3}
   \frac{1}{\nu-i\omega} \nonumber \\&\times &
    \int d^3\upsilon\,\upsilon^2_{\alpha}\,\delta(\upsilon^2-\upsilon^2_F)\,\Psi(q).
     \end{eqnarray}

Let us restrict ourselves only with the single MN of spheroidal shape.
 To calculate the residual integral, we pass to the spherical coordinate system with a $z$-axis directed along
the rotation axis of the spheroid (as we have done it in the previous section), and take into account that the components
of an electron velocity parallel ($\upsilon_{\|}$) and perpendicular ($\upsilon_{\bot}$) to this axis are defined as
\begin{equation}
 \label{eq 24}
  \upsilon_{\|}=\upsilon_z=\upsilon\cos\theta, \quad \upsilon_{\bot}=\sqrt{\upsilon^2_x+\upsilon^2_y}=\upsilon\sin\theta,
   \end{equation}
correspondingly, where
\[
\upsilon^2_{x\choose y}=\upsilon^2\sin^2\theta\cdot{\cos^2\varphi\choose\sin^2\varphi}.
\]
After integration in Eq.~(\ref{eq 23}) over the azimuthal angle $\varphi$ and over all electron velocities, in view
of Eqs.~(\ref{eq 16}), (\ref{eq 19}), we obtain for the $\|$- and $\bot$-components of the conductivity tensor the expressions
\begin{eqnarray}
 \label{eq 25}
  \sigma_{\|}^c&\equiv &\langle\sigma_{zz}^c({\bf r,\omega})\rangle =\frac{9}{4}\frac{n e^2}{m}\frac{1}{\nu-i\omega}
   \nonumber \\&\times &\int\limits_0^{\pi/2} \sin\theta\cos^2 \theta\;\Psi(\theta)\;d\theta|_{\upsilon=\upsilon_F},
    \end{eqnarray}
and
\begin{eqnarray}
 \label{eq 26}
  \sigma_{\bot}^c&\equiv &\langle\sigma_{xx}^c({\bf r,\omega})\rangle=
   \langle\sigma_{yy}^c({\bf r,\omega})\rangle\nonumber \\&=&\frac{9}{8}\frac{n e^2}{m}
    \frac{1}{\nu-i\omega}\int\limits_0^{\pi/2} \sin^3\theta\;\Psi(\theta)\;d\theta|_{\upsilon=\upsilon_F}.
     \end{eqnarray}
The subscript $\upsilon=\upsilon_F$ means that the electron velocity in the final expressions should be taken on the Fermi surface.
The $\Psi$-function in Eqs.~(\ref{eq 25}), (\ref{eq 26}) varies now with the angle $\theta$, because $q$ (see Eq.~(\ref{eq 22})) for
a spheroidal particle becomes dependent on the angle $\theta$, and can be determined as
\begin{equation}
 \label{eq 27}
  q=\frac{2}{\upsilon_F}
   \frac{\nu-i\omega}{\sqrt{\frac{\cos^2\theta}{R^2_{\|}}+\frac{\sin^2\theta}{R^2_{\bot}}}}\equiv q(\theta),
    \end{equation}
where $R_{\|}$ and $R_{\bot}$ are the semiaxes of the spheroid.
Such form for $q$ follows from the form of "deformed" electron velocity, which enters
into Eq.~(\ref{eq 22}), and for a spheroid is
\begin{eqnarray}
 \label{eq 28}
  \upsilon'=\upsilon R\sqrt{\left(\frac{\sin\theta}{R_{\bot}}\right)^2+\left(\frac{\cos\theta}{R_{\|}}\right)^2}\equiv\upsilon'(\theta),
   \end{eqnarray}
where the velocity components $\upsilon_{\|}$ and $\upsilon_{\bot}$, presented by Eq.~(\ref{eq 24}), were used.
The velocity $\upsilon'$ does not depend on the particle radius $R$, but only on the spheroid aspect ratio.
In the case of a spherical particle $R_{\|}=R_{\bot}\equiv R$, and $\upsilon'= \upsilon$.

There is a well known common relation \cite{Ruch} between the tensor components of the complex
dielectric permeability and the components of a complex conductivity tensor
\begin{equation}
 \label{eq 29}
  \langle\epsilon_{\alpha\beta}({\bf r},\omega)\rangle=
   \delta_{\alpha\beta}+i\frac{4\pi}{\omega}\langle\sigma_{\alpha\beta}^c({\bf r,\omega})\rangle.
    \end{equation}
If one separates the real and the imaginary parts in the expressions for both
the complex tensors of the dielectric permeability and the conductivity, i.e., presents
\begin{equation}
 \label{eq 30}
  \langle\epsilon_{\alpha\beta}^c({\bf r,\omega})\rangle =\epsilon'_{\alpha\beta}(\omega)+i\epsilon''_{\alpha\beta}(\omega),
   \end{equation}
and
\begin{equation}
 \label{eq 31}
  \langle\sigma_{\alpha\beta}^c({\bf r,\omega})\rangle =\sigma'_{\alpha\beta}(\omega)+i\sigma''_{\alpha\beta}(\omega),
   \end{equation}
then in a correspondence with Eq.~(\ref{eq 29}), one obtains the next two relations
\begin{equation}
 \label{eq 32}
  \epsilon'_{\alpha\beta}(\omega)=\delta_{\alpha\beta}(\omega)-\frac{4\pi}{\omega}\sigma''_{\alpha\beta}(\omega),
   \end{equation}
and
\begin{equation}
 \label{eq 33}
  \epsilon''_{\alpha\beta}(\omega)=\frac{4\pi}{\omega}\sigma'_{\alpha\beta}(\omega).
   \end{equation}
Finally, using Eqs.~(\ref{eq 25}), (\ref{eq 26}) and Eq.~(\ref{eq 31}), one gets for a spheroidal particle
\begin{equation}
 \label{eq 34}
  \sigma'_{\|\choose\bot}(\omega)=\frac{9}{4}\frac{ne^2}{m}\Re\left[\frac{1}{\nu-i\omega}\int\limits_0^{\pi/2}
   {\sin\theta\,\cos^2\theta \choose \frac{1}{2}\sin^3\theta}\Psi(\theta)\;d\theta\right]_{\upsilon=\upsilon_F}
    \end{equation}
and
\begin{equation}
 \label{eq 35}
  \sigma''_{\|\choose\bot}(\omega)=\frac{9}{4}\frac{ne^2}{m}\Im\left[\frac{1}{\nu-i\omega}\int\limits_0^{\pi/2}
   {\sin\theta\cos^2\theta \choose \frac{1}{2}\sin^3\theta}\Psi(\theta)\;d\theta\right]_{\upsilon=\upsilon_F}.
    \end{equation}
The upper (lower) symbols in the parentheses of the left hand sides of Eqs.~(\ref{eq 34}), (\ref{eq 35}) correspond to the upper
(lower) symbols in the parentheses of the right hand sides of these equations. Formulas (\ref{eq 34}), (\ref{eq 35}) are the
fundamental equations for calculations of the dielectric function and, therefore, for studying the optical properties of MNs.

For illustration, we present in Fig.~1 the frequency dependence of an imaginary part of the dielectric permeability
ratio components for a spheroidal Au particle, which is obtained  by numerical evaluating the integrals in Eqs.~(\ref{eq 33}), (\ref{eq 34}). It is worth to note that the radius $R_{\|}$ is directed along the revolution axis of the spheroid, and $R_{\bot}$ -- transverse
to it. These spheroidal radiuses can be expressed through the radius of a sphere (of an equivavalent volume) as
$$R=(R_{\|} R^2_{\bot})^{1/3}.$$
The calculations were carried out using such parameters for Au:
$\nu=3.39\times 10^{13}$ at $0^0 C$ [\onlinecite{GT}], $n=5.9\times 10^{22} cm^{-3}$,
$\upsilon_F=1.39\times 10^8 cm/s$ [\onlinecite{CK}].

\vskip10pt
\noindent\includegraphics[width=8.6cm]{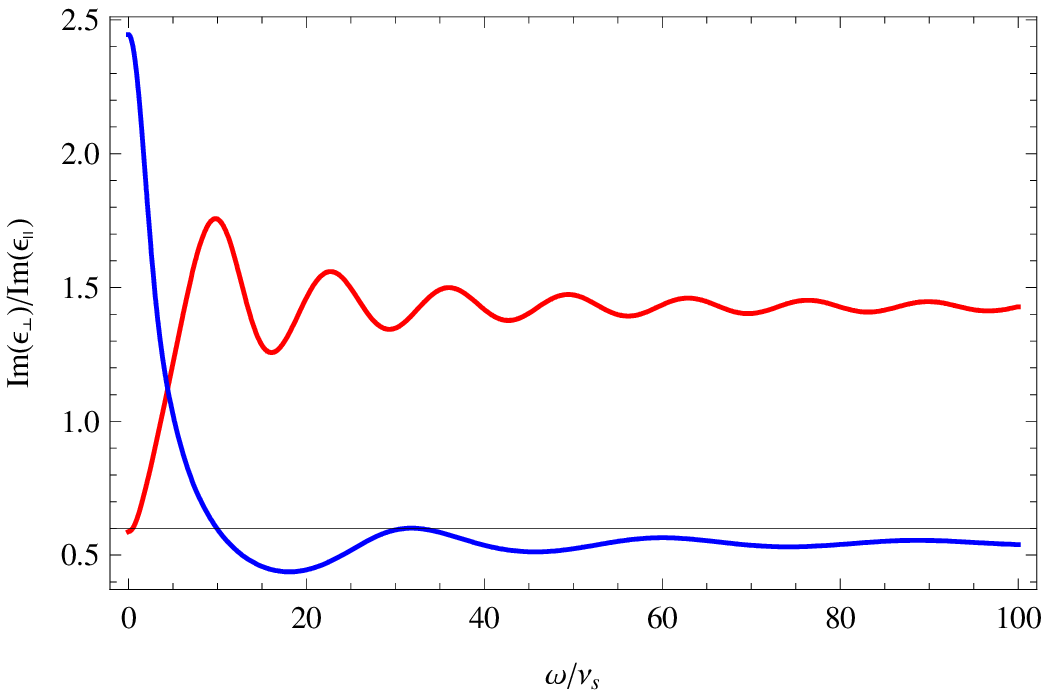}
\vskip-1mm\noindent{\footnotesize FIG.~1. The dependence of the ratio $\epsilon''_{\bot}/\epsilon''_{\|}$
for prolate ($R_{\bot}/R_{\|}=0.1$, upper curve) and oblate ($R_{\bot}/R_{\|}=10$, lower curve) Au particle
(with $R=50 \AA$) versus frequencies ratio $\omega/\nu_s$, where $\nu_s=\upsilon_F/(2R)$.}
\vskip10pt

As one can see, the ratio of $\epsilon''_{\bot}/\epsilon''_{\|}$ oscillates with increasing frequency both for the prolate and the oblate Au nanoparticles. These oscillations have a damping character both for prolate and oblate particles, but differ in the period of oscillations. For a given prolate nanoparticle the oscillations occur around the constant value $\epsilon''_{\bot}/\epsilon''_{\|}\simeq 4/3$ and for an oblate ones -- in the vicinity of the another constant $\epsilon''_{\bot}/\epsilon''_{\|}\simeq 1/2$. The period of oscillations depend only slightly on the particle volume, however, the oscillation amplitude is more pronounced for particles of smaller radii.

It is also worth noting that the intensity of the surface mode is determined by the magnitude of the imaginary component of the material dielectric constant. Materials with a small $\epsilon''$ have a large, narrow absorption peak, whereas materials with a large $\epsilon''$ have a small broader absorption peak.\cite{HFM}

The ratio of the real parts of $\epsilon'_{\bot}/\epsilon'_{\|}$ does not oscillate with frequency, therewith at the frequency
of $\omega\simeq\omega_{pl}$ exhibits the singularity. Thus, we demonstrate below the plots for $\epsilon'_{\bot}(\omega)$
and $\epsilon'_{\|}(\omega)$ separately.

Figure 2 shows the real part of the dielectric permeability components for Au nanoparticle as a function
of normalized frequency, obtained by using Eqs.~(\ref{eq 32}), (\ref{eq 35}) in numerical calculations. The magnitude of $\epsilon'$
reaches minimum value at $\omega\rightarrow 0$, and $\epsilon'\rightarrow 0$ at $\omega\rightarrow\omega_{pl}$. Among the two components
of $\epsilon'$, the frequency dependence is more pronounced for the longitudinal one: $\epsilon'_{\|}$ attains the smallest negative
value at $\omega\rightarrow 0$ in the case of the prolate Au nanoparticle, and the maximal negative value -- for the oblate one.
The absolute magnitude of the both components of $\epsilon'$ is essentially enhanced as the radius of the particle is increased
(especially, at $\omega\rightarrow 0$).

\vskip10pt
\noindent\includegraphics[width=8.6cm]{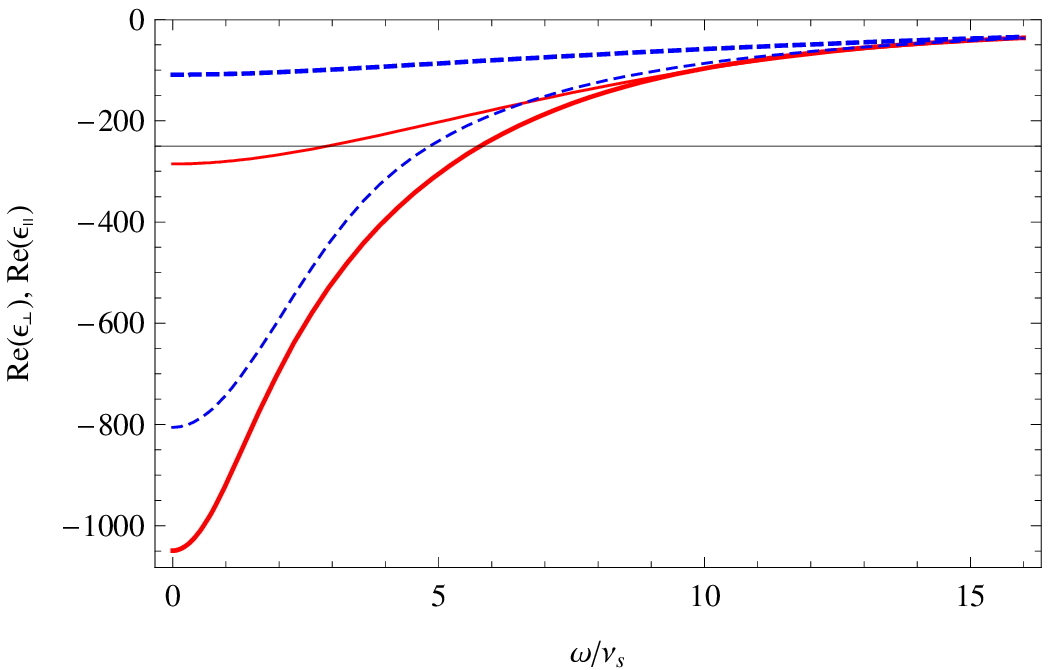}
\vskip-1mm\noindent{\footnotesize FIG.~2. The dependence of $\epsilon'_{\bot}, \epsilon'_{\|}$
for both the prolate ($R_{\bot}/R_{\|}=0.1$, solid lines) and the oblate ($R_{\bot}/R_{\|}=10$,
dashed lines) Au particle (with $R=50 \AA$) versus frequencies ratio $\omega/\nu_s$. Thick curves
are for $\|$-components, and thin curves -- for $\bot$-components of $\epsilon'$.}
\vskip10pt

Below, we will consider several approaches, enable us to derive the explicit analytical expressions
for $\sigma'$ and $\sigma''$ from Eqs.~(\ref{eq 34}), (\ref{eq 35}).

\section{LIMIT CASES. SIZE EFFECTS}
\subsection{Frequency approach}

The straightforward evaluations of the dielectric permeability of a single MN or it conductance can be made, when
$\Psi$-function entering in Eq.~(\ref{eq 23}) or in Eqs.~(\ref{eq 34}), (\ref{eq 35}) takes the simplest form.
Let us introduce the value
\begin{equation}
 \label{eq 36}
  \nu_s=\frac{\upsilon'}{2R},
   \end{equation}
which will characterize the frequency of electron collisions with the spherical MN surfaces,
and $R$ is the sphere radius. This allow us to rewrite Eq.~(\ref{eq 22}) as
\begin{equation}
 \label{eq 37}
  q=\frac{\nu}{\nu_s}-i\frac{\omega}{\nu_s}.
   \end{equation}

i) First, we consider, for example, the case $|q|\gg 1$. In the frequency scale, this implies that both
inequalities $\nu\gg\nu_s$ and $\omega\gg\nu_s$ must be executed. Then, Eq.~(\ref{eq 21}) reduces to the form
\begin{equation}
 \label{eq 38}
  \Psi(q)|_{q\gg 1}\approx \frac{4}{3}-\frac{2}{q}+\frac{4}{q^3}-\cdots.
   \end{equation}
Accounting for Eqs.~(\ref{eq 38}) and (\ref{eq 22}), we can calculate approximately the real and imaginary parts of the ratio
\begin{equation}
 \label{eq 39}
  \Re\left(\frac{\Psi(\theta)}{\nu-i\omega}\right)_{|q|\gg 1}\approx\frac{4}{3}\frac{\nu}{\nu^2+\omega^2}-
   \frac{\upsilon'(\theta)}{R}\frac{\nu^2-\omega^2}{(\nu^2+\omega^2)^2}+\cdots,
    \end{equation}
\begin{equation}
 \label{eq 40}
  \Im\left(\frac{\Psi(\theta)}{\nu-i\omega}\right)_{|q|\gg 1}\approx\frac{4}{3}\frac{\omega}{\nu^2+\omega^2}-2
   \frac{\upsilon'(\theta)}{R}\frac{\nu\;\omega}{(\nu^2+\omega^2)^2}+\cdots,
    \end{equation}
which enters into Eqs.~(\ref{eq 34}), (\ref{eq 35}). Here, the velocity $\upsilon'(\theta)$ is given by Eq.~(\ref{eq 28}).
The formula (\ref{eq 39}) at $\nu\rightarrow 0$ agrees with an earlier estimation given in Ref. [\onlinecite{GT1}].

Substituting Eq.~(\ref{eq 39}) into Eq.~(\ref{eq 34}), and Eq.~(\ref{eq 40}) into Eq.~(\ref{eq 35}), correspondingly,
and using the values of the integrals $I_{\|}$ and $I_{\bot}$ given in the Appendix A, one obtains for the real
and imaginary parts of the conductivity tensor components the expressions
\begin{equation}
 \label{eq 41}
  \sigma'_{\|\choose\bot}(\omega)\approx\frac{ne^2}{m}\left(\frac{\nu}{\nu^2+\omega^2}-
   \frac{9}{2}\nu_s\frac{\nu^2-\omega^2}{(\nu^2+\omega^2)^2}I_{\|\choose\bot}+\cdots\right),
    \end{equation}
\begin{equation}
 \label{eq 42}
  \sigma''_{\|\choose\bot}(\omega)\approx\frac{ne^2}{m}\left(\frac{\omega}{\nu^2+\omega^2}-
   9\nu_s\frac{\nu\;\omega}{(\nu^2+\omega^2)^2}I_{\|\choose\bot}+\cdots\right),
    \end{equation}
provided that the conditions $\nu\gg\nu_s$, $\omega\gg\nu_s$ and $\upsilon=\upsilon_F$ are satisfied.
To meet the requirement of $\nu\gg\nu_s$, e.g., for Au particle at $0^o C$, it is necessary that its radius should
be $R\gg 400\AA$. The first term in the parentheses of Eqs.~(\ref{eq 41}), (\ref{eq 42}) results from the integration over angle
\begin{equation}
 \label{eq 43}
  \int\limits_0^{\pi/2}{\sin\theta\,\cos^2\theta \choose \frac{1}{2}\sin^3\theta}\,d\theta =\frac{1}{3}.
   \end{equation}
The second term in the parentheses of Eqs.~(\ref{eq 41}), (\ref{eq 42}) contains the integrals $I$ which accounts for the particle nonsphericity and doesn't depend on the frequency. The simplest result for $\sigma$ can be obtained for a spherical MN, when the
integrals $I$ for different polarizations coincide with each other, and are equal to $1/3$ (see Eqs.~(\ref{eq 90}) in Appendix A).

 As one can see, the first term in the parentheses of Eqs.~(\ref{eq 41}), (\ref{eq 42}) describes the Drude-Sommerfeld results
 for a spherical particle, and the second one gives the first correction of the kinetic theory to the volume electron scattering,
 allowing to account an electron scattering from the surfaces of the nanoparticle as well. The calculations of the frequency dependencies
 of the ratio $\epsilon''_{\bot}/\epsilon''_{\|}$ with employing of Eqs.~(\ref{eq 33}) and (\ref{eq 41}) (and the conditions pointed therein) give the results qualitatively similar to our previous numerical calculations (using Eqs.~(\ref{eq 33}), (\ref{eq 34})), depicted in Fig.~1, though without any oscillations. But quantitatively, the results obtained from Eq.~(\ref{eq 34}) are of one or even two orders of magnitude lower (depending on the radius of MN) than those following else from the Drude-Sommerfeld formula or from Eqs.~(\ref{eq 41}).

ii) For the opposite limit case, when $|q|\ll 1$ (or $\nu\ll\nu_s$, and $\omega\ll\nu_s$),
we can take advantage of the expansion
\[
e^{-q}\simeq 1-q+\frac{q^2}{2!}-\frac{q^3}{3!}+\frac{q^4}{4!}-\frac{q^5}{5!}+\frac{q^6}{6!}-\cdots,
\]
and then from Eq.~(\ref{eq 21}), we immediately obtain
\begin{equation}
 \label{eq 44}
  \Psi(q)|_{|q|\ll 1}\approx \frac{1}{2}q-\frac{2}{15}q^2+\frac{1}{36}q^3-\cdots.
   \end{equation}
In this case, we can find for real and imaginary parts of the mentioned above ratio the relations
\begin{equation}
 \label{eq 45}
  \Re\left(\frac{\Psi(\theta)}{\nu-i\omega}\right)_{|q|\ll 1}\approx\frac{R}{\upsilon'(\theta)}-\frac{8}{15}\nu
   \left(\frac{R}{\upsilon'(\theta)}\right)^2+\cdots,
    \end{equation}
\begin{equation}
 \label{eq 46}
  \Im\left(\frac{\Psi(\theta)}{\nu-i\omega}\right)_{|q|\ll 1}\approx
   \frac{8}{15}\omega\left(\frac{R}{\upsilon'(\theta)}\right)^2+\cdots.
    \end{equation}
This permits us to get for $\sigma$ components the following expressions:
\begin{equation}
 \label{eq 47}
  \sigma'_{\|\choose\bot}\approx\frac{9}{4}\frac{ne^2}{m}\left(\frac{R}{\upsilon}I_{\|\choose\bot}^{<}|_{\upsilon=
   \upsilon_F}-\frac{2}{15}\frac{\nu}{\nu^2_s} J_{\|\choose\bot}+\cdots\right),
    \end{equation}
\begin{equation}
 \label{eq 48}
  \sigma''_{\|\choose\bot}(\omega)\approx\frac{9}{30}\frac{ne^2}{m}
   \frac{\omega}{\nu^2_s} \left(J_{\|\choose\bot}+\cdots\right),
    \end{equation}
provided that $\nu\ll\nu_s$, $\omega\ll\nu_s$, and $\upsilon=\upsilon_F$. In other words, the first condition means that the radius of MN should obey the condition $R\ll \upsilon_F/(2\nu)$. The first term in the parentheses of Eqs.~(\ref{eq 47}) is the result of the integration over angle $\theta$ with $\upsilon'(\theta)$ in the denominator (see Appendix A, Eqs.~(\ref{eq 91}), (\ref{eq 92})), and the second one can be given by
\begin{eqnarray}
 \label{eq 49}
  J_{\|\choose\bot}&=&\frac{1}{R^2}\frac{R_{\|}^2 R_{\bot}^2}{R_{\bot}^2-R_{\|}^2}
   \left[\pm {1\choose 1/2}\mp {1\choose \frac{1}{2}(R_{\bot}/R_{\|})^2}\right.
    \nonumber \\&\times&\left.\frac{R_{\|}}{\sqrt{R_{\bot}^2-R_{\|}^2}}
     \arctan{\frac{\sqrt{R_{\bot}^2-R_{\|}^2}}{R_{\|}}}\right].
      \end{eqnarray}
The upper (lower) signs and upper (lower) symbols in the parentheses of the right hand side of Eq.~(\ref{eq 49})
correspond to those in the left hand side of this equation. In the case of prolate spheroidal particles ($R_{\|}>R_{\bot}$),
one should made in Eqs.~(\ref{eq 49}) only the following replacement
\begin{equation}
 \label{eq 50}
  \arctan{\frac{\sqrt{R_{\bot}^2-R_{\|}^2}}{R_{\|}}}
   \rightarrow \frac{1}{2i}\ln\left|\frac{R_{\|}-\sqrt{R^2_{\|}-R^2_{\bot}}}{R_{\|}+\sqrt{R_{\|}^2-{R_{\bot}^2}}}\right|,
    \end{equation}
and in the case of a spherical symmetric MN, $R_{\|}=R_{\bot}\equiv R$, and $J_{\|}=J_{\bot}\equiv 1/3$.

As one can see from Eq.~(\ref{eq 47}), the real part of $\sigma_{\|\choose\bot}$ doesn't depend on the frequency,
but only on the particle geometry, which is defined here by the parameters $I^<$ and $J$. For imaginary part of
$\sigma_{\|\choose\bot}$, we have from Eq.~(\ref{eq 48}) the linear enhancement with the frequency. This implies that
\begin{equation}
 \label{eq 51}
  \epsilon'_{\|\choose\bot}\approx 1-\frac{9}{30} \left(\frac{\omega_{pl}}{\nu_s}\right)^2 J_{\|\choose\bot},
    \end{equation}
doesn't depend on the frequency, provided that the conditions $\omega\ll\nu_s$ and $\nu\ll\nu_s$ are fulfilled.

\subsection{Size approach}

It remains to examine the cases of different relations between the particle sizes and the conduction electron
mean-free path inside a particle. Using Eq.~(\ref{eq 28}), we can rewrite Eq.~(\ref{eq 22}) in somewhat another form
\begin{equation}
 \label{eq 52}
  q=q_1(\theta)-iq_2(\theta),
   \end{equation}
with
\begin{equation}
 \label{eq 53}
  q_1(\theta)=\frac{1}{\sqrt{\left(\frac{l}{2R_{\|}}\right)^2\cos^2\theta+
   \left(\frac{l}{2R_{\bot}}\right)^2\sin^2\theta}},
    \end{equation}
where $2R_{\|}$ is the length of the MN along the $z$ axis, which is directed along the
principal spheroid axis, $2R_{\bot}$ is the MN size along the $x$ or $y$ directions, and
\begin{equation}
 \label{eq 54}
  l=\frac{\upsilon_F}{\nu}
   \end{equation}
has the sense of the length of an electron mean-free path (for Au at $0^0 C$, e.g., $l\simeq 410 \AA$), and
\begin{equation}
 \label{eq 55}
  q_2(\theta)=\frac{1}{\sqrt{\left(\frac{\nu_{s \bot}}{\omega}\right)^2\sin^2\theta+
   \left(\frac{\nu_{s \|}}{\omega}\right)^2\cos^2\theta}}.
    \end{equation}
Here
\begin{equation}
 \label{eq 56}
  \nu_{s \|}=\frac{\upsilon_F}{2R_{\|}},  \qquad  \nu_{s \bot}=\frac{\upsilon_F}{2R_{\bot}}
   \end{equation}
are the frequencies of the electron collision with the particle surfaces along and across the $z$-axis of a spheroid, correspondingly.

Let us consider the possible relations between $l$ and particle sizes $2R_{\|}$, $2R_{\bot}$.

i) The conduction electron mean-free path is much less than the sizes of the particle along particular directions: $l\ll 2R_{\bot}$,
\quad  $l\ll 2R_{\|}$. As follows from Eq.~(\ref{eq 53}), $q_1(\theta)|_{\upsilon=\upsilon_F}\gg 1$. In this case, an electron is
scattered predominately inside the volume of MN. If, moreover, $q_2\rightarrow 0$, i.e., $\nu_{s,\|}, \nu_{s,\bot}\gg\omega$, then from Eq.~(\ref{eq 21}), one gets
\begin{equation}
 \label{eq 57}
  \Psi\simeq \frac{4}{3}.
   \end{equation}
Substituting Eq.~(\ref{eq 57}) into Eqs.~(\ref{eq 34}), (\ref{eq 35}), and using Eq.~(\ref{eq 43}), we obtain
the Drude-Sommerfeld formulas for real and imaginary parts of $\sigma$, presented above by the first term in
the parenthesis of Eqs.~(\ref{eq 41}), (\ref{eq 42}).

ii) The mean-free path of a conduction electron is much greater than the particle size along particular directions:
$l\gg 2R_{\bot}$, \quad  $l\gg 2R_{\|}$. In this case, an electron scattering occurs mainly from the inner surface of the MN.
The electrons oscillate between the walls of the particle with different frequencies, excluding $\nu_{s,\|}$,
$\nu_{s,\bot}$. In accordance with Eq.~(\ref{eq 53}), the inequality $q_1(\theta)\ll 1$ holds only for $q_1$.
The parameter $q_2$ remains arbitrary. Formally, we can put $q_1(\theta)\rightarrow 0$, and for a real and imaginary
parts of $\Psi$-function one finds
\begin{equation}
 \label{eq 58}
  \Re\;\Psi(q)|_{q_1\rightarrow 0}=\frac{4}{3}+\frac{4}{q^2_2}\left(\cos q_2 -\frac{1}{q_2}\sin q_2\right),
   \end{equation}
and
\begin{equation}
 \label{eq 59}
  \Im\;\Psi(q)|_{q_1\rightarrow 0}=-\frac{2}{q_2}-\frac{4}{q^3_2}+\frac{4}{q^2_2}\left(\frac{1}{q_2}\cos q_2 +\sin q_2\right).
   \end{equation}
The equations (\ref{eq 58}), (\ref{eq 59}) allow to fulfil the calculation of the real and
the imaginary parts of the ratio $\Psi/(\nu-i\omega)$, as we have done it before. Then, we obtain

\begin{equation}
 \label{eq 60}
  \Re\left[\frac{\Psi(q)}{\nu-i\omega}\right]_{q_1\rightarrow 0}=
   \frac{1}{\omega}\left[\frac{2}{q_2}-\frac{4}{q^2_2}\sin q_2+\frac{4}{q_2^3}(1-\cos q_2)\right],
    \end{equation}
\begin{equation}
 \label{eq 61}
  \Im\left[\frac{\Psi(q)}{\nu-i\omega}\right]_{q_1\rightarrow 0}=\frac{4}{\omega}\left[\frac{1}{3}+
   \frac{1}{q^2_2}\left(\cos q_2-\frac{1}{q_2}\sin q_2\right)\right].
    \end{equation}
The parameter $q_2$, defined by Eq.~(\ref{eq 55}), is governed by the frequency. Depending on the ratio between
the incident frequency and the frequencies $\nu_{s,\|}$, $\nu_{s,\bot}$, the value of $q_2$ can be greater or less than 1.
This makes it difficult to perform subsequent analytical calculations of the integrals involved into Eqs.~(\ref{eq 34}),
(\ref{eq 35}). Below, we dwell on some particular cases for which the calculations of $\sigma$ are the most simple. It should
be noted also that the corresponding expressions for the components of $\epsilon$ can be easily obtained by substituting
of $\sigma$ into Eqs.~(\ref{eq 32}), (\ref{eq 33}).

\subsubsection{Conductivity of a spherical MN}

In the case of a spherical MN there are three characteristic frequencies which are considered usually: the frequency of an incident electromagnetic field $\omega$, the collision frequency of electrons in the particle volume $\nu$, and the vibration frequency between
the particle walls $\nu_s$ (if the particle size is less than the electron mean free path). When $\nu>\nu_s$, the mechanism of an electron scattering in the bulk is dominated, and an electron scattering from the particle surface gives only small corrections of the order of $\nu_s/\nu$. But we are interested in the case, when the mechanism of the surface electron scattering dominates, which corresponds to $\nu<\nu_s$.

For particles of a spherical shape, the electric conductivity becomes a scalar quantity, and one can put $R_{\|}=R_{\bot}\equiv R$
in Eq.~(\ref{eq 27}), then $q$ and $\Psi$-function in Eqs.~(\ref{eq 25}), (\ref{eq 26}) are not depended on the angle $\theta$.
With an account for Eq.~(\ref{eq 43}), this makes it possible to obtain for $\sigma$ the simple expression
\begin{equation}
 \label{eq 62}
  \sigma_{sph}^c=\sigma_{\|}^c=\sigma_{\bot}^c=\frac{3}{4}\frac{n e^2}{m}
   \frac{\Psi(q)}{\nu-i\omega}|_{\upsilon=\upsilon_F},
    \end{equation}
with $q=2R(\nu-i\omega)/\upsilon_F$. To calculate $\sigma_{sph}^c$, one can use either Eqs.~(\ref{eq 39}), (\ref{eq 40}),
or Eqs.~(\ref{eq 45}), (\ref{eq 46}), or Eqs.~(\ref{eq 60}), (\ref{eq 61}) for different limit cases considered above.
For example, we restrict ourselves here only to the case, when $\nu\ll\nu_s$. In this limit case, to a first approximation,
one can put $q_1\rightarrow 0$. Then Eqs.~(\ref{eq 60}), (\ref{eq 61}) with $q_2=\omega/\nu_s$ can be used in Eq.~(\ref{eq 62}).
As a result, one obtains for real and imaginary parts of $\sigma$ the following expressions
\begin{equation}
 \label{eq 63}
  \sigma'_{sph}\simeq\frac{3}{8\pi}\nu_s\frac{\omega^2_{pl}}{\omega^2}\left[1-
   \frac{2\nu_s}{\omega}\sin\frac{\omega}{\nu_s}+\frac{2\nu^2_s}{\omega^2}\left(1-
    \cos\frac{\omega}{\nu_s}\right)\right],
     \end{equation}
\begin{equation}
 \label{eq 64}
  \sigma''_{sph}\simeq\frac{\omega^2_{pl}}{4\pi\omega}\left[1+3\left(\frac{\nu_s}{\omega}\right)^2
   \left(\cos\frac{\omega}{\nu_s}-\frac{\nu_s}{\omega}\sin\frac{\omega}{\nu_s}\right)\right],
    \end{equation}
provided that $\nu\ll\nu_s$, where $\nu_s=\upsilon_F/(2R)$.
The last expression in the Drude-Sommerfeld approximation looks like\cite{BH}
\begin{equation}
 \label{eq 65}
  \sigma''_{sph}=\frac{\omega}{4\pi}\frac{\omega^2_{pl}}{\nu^2+\omega^2}.
   \end{equation}

If one puts the oscillation terms in Eqs.~(\ref{eq 63}), (\ref{eq 64}) equal to zero, and uses Eqs.~(\ref{eq 32}), (\ref{eq 33}),
then one obtains the expression for real part of the dielectric function which coincides with Eq.~(\ref{eq 1}) at $\nu\rightarrow 0$;
but for imaginary part of the dielectric function, one gets Eq.~(\ref{eq 2}) only if the replacement $\nu\rightarrow 3\nu_s/2$
will be done. This replacement is the same one as presented by Eq.~(\ref{eq 3}), which has been often used in a phenomenological approximation.

\vskip10pt
\noindent\includegraphics[width=8.6cm]{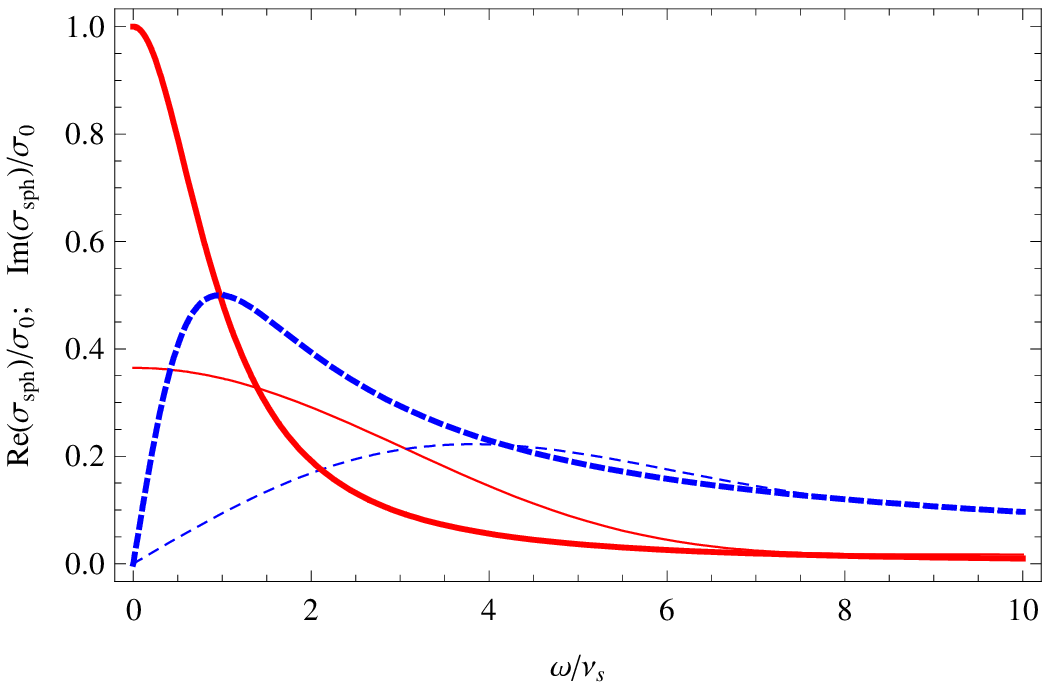}
\vskip-1mm\noindent{\footnotesize FIG.~3. The real $\sigma'$ (solid lines) and imaginary $\sigma''$ (dashed lines) parts of
the ratio of an electric conductivity to the statical conductivity $\sigma_0$ vs frequencies ratio, for a spherical Au
particle with $R=200 \AA$. The thin lines correspond to the results obtained using the kinetic method and the bold lines --
the Drude-Sommerfeld formulas.}
\vskip10pt

The calculated results of the real and imaginary parts of an electric conductivity across the normalized to $\nu_s$ frequency are shown in Fig.~3 for spherical Au nanoparticle, obtained with the use of both the kinetic method and the Drude-Sommerfeld formulas. The conductivity is measured on the scale of the statical conductivity $\sigma_0=ne^2/(m\nu)$. For calculations of $\sigma$, Eqs.~(\ref{eq 34}), (\ref{eq 35}) have been used in the kinetic case, and Eqs.~(\ref{eq 2}) and (\ref{eq 65}) -- in the Drude-Zommerfeld case. For illustration, the numerical parameters for Au particle\cite{CK} and $\omega_{pl}=1.37\times 10^{16} s^{-1}$ were chosen. As it can be seen in Fig.~3,
the kinetic method appreciably changes the frequency dependence of $\sigma_{sph}$ at low frequencies, and at high frequencies ($\omega\gg\nu_s$) gives the same result for $\sigma_{sph}$ as that which follows from the Drude-Zommerfeld formula. The difference between kinetic and Drude-Zommerfeld results is enhanced markedly as the particle radius is decreased. The real part of $\sigma$ is peaked at $\omega/\nu_s\rightarrow 0$ in both cases, whereas the imaginary part of $\sigma$ -- at $\omega=\nu_s$ in the Drude-Zommerfeld case, and at $\omega\approx 4\nu_s$ -- using the kinetic method.

For an extremely low $\omega\ll\nu_s$ or an extremely high $\omega\gg\nu_s$ frequencies, one can find
from Eqs.~(\ref{eq 63}), (\ref{eq 64}) after some algebra, the next simple approximation for $\sigma'$
\begin{equation}
 \label{eq 66}
  \sigma'_{sph}=\frac{3}{16\pi}\omega^2_{pl}\cdot\left\{{\frac{R}{\upsilon_F},\;\;\;
   \omega\ll\nu_{s},\atop\frac{\upsilon_F}{R\omega^2},\;\;\;\omega\gg\nu_{s}}\right.,
    \end{equation}
and for $\sigma''$:
\begin{equation}
 \label{eq 67}
  \sigma''_{sph}=\frac{\omega^2_{pl}}{4\pi}
   \cdot\left\{ \frac{\omega}{2}\left(\frac{R}{\upsilon_F}\right)^2,\;\;\;
    \omega\ll\nu_{s},\atop\frac{1}{\omega},\;\;\;\omega\gg\nu_{s}\right..
     \end{equation}
The results of Eqs.~(\ref{eq 66}), (\ref{eq 67}) at $\omega\gg\nu_{s}$ correspond to those presented
above by first terms in the square brackets of Eq.~(\ref{eq 63}), (\ref{eq 64}), accordingly.

The calculations with employing of Eqs.~(\ref{eq 63}), (\ref{eq 64}) for Au particle with $R=200\AA$ give the results
similar qualitatively to the ones above presented, but quantitatively they are at the $\sigma$ maximum of approximately 30\% higher.

In spite of the oscillation terms in square brackets of Eqs.~(\ref{eq 63}), (\ref{eq 64}), the ratio of both $\sigma_{sph}'/\sigma_0$
and $\sigma_{sph}''/\sigma_0$ does't oscillate with the frequency owing to the cut-off factors before these brackets.

\vskip10pt
\noindent\includegraphics[width=8.6cm]{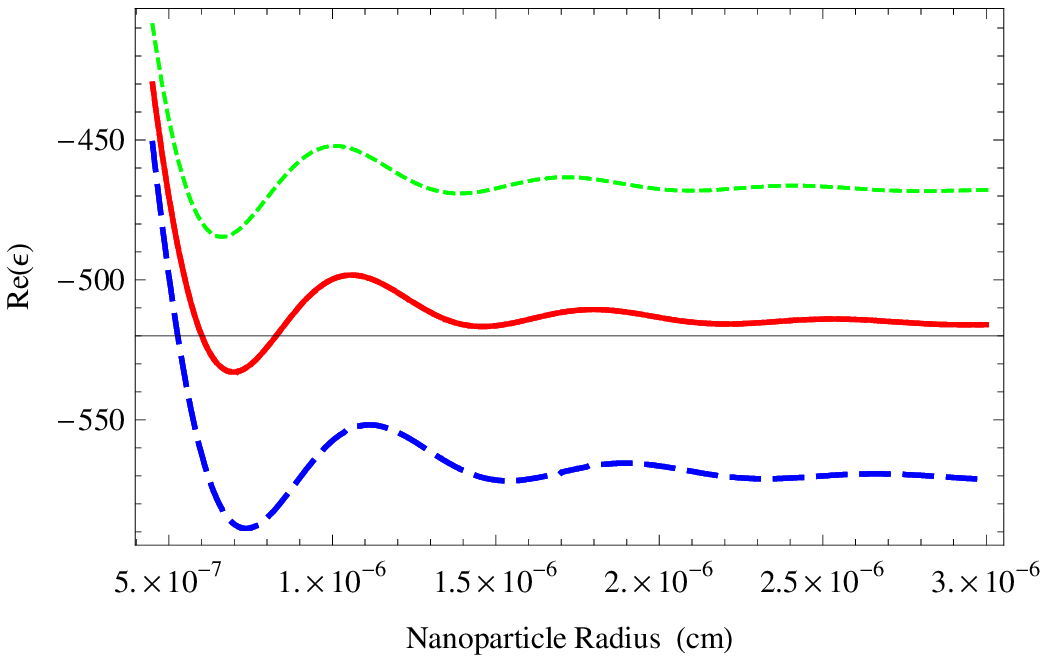}
\vskip-1mm\noindent{\footnotesize FIG.~4. The dependence of $\epsilon'$
for the spherical Au particle versus a radius $R$ at the frequencies of $\omega=$: $5.7\times 10^{14} s^{-1}$ (dashed line),
$6\times 10^{14} s^{-1}$ (solid line), and $6.3\times 10^{14} s^{-1}$ (doted line)}.
\vskip10pt

Now, let us discuss shortly the dependence of the dielectric permeability on the size of MN. In Fig.~4, we present
the results of our numerical calculations of $\epsilon(R)$ using Eqs.~(\ref{eq 32}), (\ref{eq 35}) at the fixed $\omega$.
For illustration, we choose such frequencies from the frequency scale for which the above dependencies are the most
pronounced for Au particle.

\vskip10pt
\noindent\includegraphics[width=8.6cm]{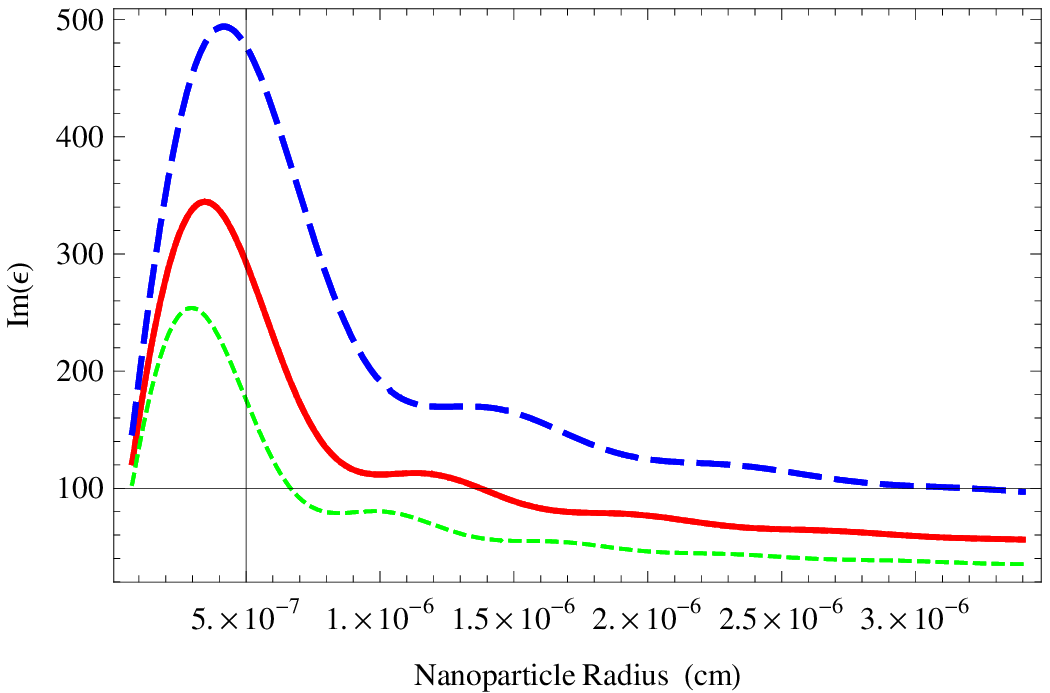}
\vskip-1mm\noindent{\footnotesize FIG.~5. The dependence of $\epsilon''$
for the spherical Au particle across the radius $R$ at the frequencies of $\omega=$:
$5\times 10^{14} s^{-1}$ (dashed line), $6\times 10^{14} s^{-1}$ (solid line), and $7\times 10^{14} s^{-1}$ (doted line).}
\vskip10pt

The real $\epsilon'$ as well as the imaginary $\epsilon''$ parts of the dielectric permeability oscillate,
when the particle radius is increased. These oscillations have a damping character and practically are vanished for
MNs of high radiuses. The real part of $\epsilon$ tends to $+1$, and the imaginary part tends to $0$ at $\omega\rightarrow 0$.
The real part has a first main minimum and the imaginary part has a first main maximum at small values of $R$.
Both the minimum and the maximum of $\epsilon$ are slightly shifted towards the greater $R$ with the frequency decreasing.
In other words, this means that the resonance energy peak shifts toward the red-side with increasing sizes of Au particle.
The oscillation period in accordance with Eqs.~(\ref{eq 63}), (\ref{eq 64}) is defined by
\begin{equation}
 \label{eq 68}
  T=\pi\frac{\upsilon_F}{R\omega}
   \end{equation}
 and, as one can see, essentially depends on the product of $R\omega$. It becomes shorter at $R\omega\ll \upsilon_F$,
 and at $R\omega\gg \upsilon_F$ -- extends. Two types of oscillations may occur: at fixed $\omega$ with
 varying of $R$, or at fixed $R$ with changing of $\omega$. For every fixed frequency there is the constant
 of $\epsilon'$ or $\epsilon''$ around which those quantities oscillate with altering of $R$. At high frequencies,
 the period of oscillations is sharply decreased and their amplitude are considerably lowered proportionally
 to the factors $(\omega_{pl}/\omega)^2$ or $\omega_{pl}^2/\omega$ before square brackets in Eqs.~(\ref{eq 63}),
 (\ref{eq 64}), correspondingly.

 If we assume that the dielectric function of a surrounding media $\epsilon_m$ is close to unity, then for a spherical
 Au particle at a plasmon frequency $\omega=\omega_{pl}/\sqrt{1+2\epsilon_m}$ we get the closely packed oscillations
 of $\epsilon'(R)$ within the amplitude interval $-2\div 1$.
 The behavior of $\sigma'(R)$ and $\sigma''(R)$ at the frequencies $\omega\ll \nu_s$ and $\omega\gg \nu_s$ can be
 seen from Eqs.~(\ref{eq 66}), (\ref{eq 67}) as well.

\subsubsection{Conductivity of an oblate MN}

For oblate particles
\begin{equation}
 \label{eq 69}
  l>2R_{\bot}>2R_{\|}.
   \end{equation}
We shall consider, for convenience, the frequency intervals $\omega\ll\nu_{s,\bot}$ and $\omega\gg\nu_{s,\|}$.
According to Eq.~(\ref{eq 55}), the former interval corresponds to $q_{2}|_{\upsilon=\upsilon_F}<1$, and the
latter one -- to $q_{2}|_{\upsilon=\upsilon_F}>1$, respectively.
It is easy to show that Eqs.~(\ref{eq 60}), (\ref{eq 61}) for these frequency intervals transform to
\begin{equation}
 \label{eq 70}
  \Re\left[\frac{\Psi(q)}{\nu-i\omega}\right]_{q_1\rightarrow 0}\approx
   \frac{1}{\omega}\cdot\left\{ {\frac{q_2}{2},\;\;\;\omega\ll\nu_{s,\bot},\atop\frac{2}{q_2},\;\;\;\omega\gg\nu_{s,\|}}\right.,
    \end{equation}
\begin{equation}
 \label{eq 71}
  \Im\left[\frac{\Psi(q)}{\nu-i\omega}\right]_{q_1\rightarrow 0}\approx
   \frac{1}{\omega}\cdot\left\{ {\frac{q_2^2}{6},\;\;\;\omega\ll\nu_{s,\bot},\atop\frac{4}{3},\;\;\;\omega\gg\nu_{s,\|}}\right..
    \end{equation}
Then, the integrals in Eqs.~(\ref{eq 34}), (\ref{eq 35}) can be calculated exactly in the approximations of
(\ref{eq 70}), (\ref{eq 71}), and for an arbitrary spheroid aspect ratio between $R_{\|}$ and $R_{\bot}$
we obtain the following expressions for the components of an electric conductivity
\begin{equation}
 \label{eq 72}
  \sigma'_{\|\choose\bot}(\omega)\simeq\frac{9}{16\pi}\frac{\omega^2_{pl}}{\omega}\cdot
   \left\{ \frac{R \omega}{\upsilon_F} I^<_{\|\choose\bot},\;\;\;\omega\ll\nu_{s,\bot}
    \atop\frac{\upsilon_F}{R\omega} I^>_{\|\choose\bot},\;\;\;\omega\gg\nu_{s,\|}\right.,
     \end{equation}
\begin{equation}
 \label{eq 73}
  \sigma''_{\|\choose\bot}(\omega)\simeq\frac{1}{4\pi}\frac{\omega^2_{pl}}{\omega}\cdot\left\{
   \frac{3}{2}\left(\frac{R\omega}{\upsilon_F}\right)^2 J_{\|\choose\bot},
    \;\;\;\omega\ll\nu_{s,\bot} \atop 1,\;\;\;\omega\gg\nu_{s,\|}\right..
     \end{equation}
In the case of $\omega\ll\nu_{s,\bot}$, Eq.~(\ref{eq 72}) coincides with the first term found previously in
Eq.~(\ref{eq 47}), and in the case of $\omega\gg\nu_{s,\|}$, it coincides with the second term of Eq.~(\ref{eq 41})
at $\nu=0$. Similarly, Eq.~(\ref{eq 73}) in the case of $\omega\ll\nu_{s,\bot}$ agrees with the first term found
previously in Eq.~(\ref{eq 48}) with an accuracy of a numerical coefficient, and in the case of $\omega\gg\nu_{s,\|}$
coincides with the first term of Eq.~(\ref{eq 42}) at $\nu=0$.

The above expressions can be transformed to their simplest forms for strongly deformed particles. Thus, for strongly oblate MNs, when
$R_{\bot}\gg R_{\|}$, using Eqs.~(\ref{eq 88}), (\ref{eq 94}) from Appendix A, it can be easily found that at low frequencies
\begin{equation}
 \label{eq 74}
  \left.\sigma'_{\|}=\frac{9}{32\pi}\omega^2_{pl}\frac{R_{\|}}{\upsilon_F},
   \atop
    \sigma'_{\bot}=\sigma'_{\|}\left(\ln{2\frac{R_{\bot}}{R_{\|}}}-\frac{1}{2}\right)\right\},\;\;\;\omega\ll\nu_{s,\bot},
     \end{equation}
and at high frequencies:
\begin{equation}
 \label{eq 75}
  \left.\sigma'_{\|}=\frac{9}{64\pi}\left(\frac{\omega_{pl}}{\omega}\right)^2\frac{{\upsilon_F}}{R_{\|}},
   \atop
    \sigma'_{\bot}=\sigma'_{\|}/2\right\},\;\;\;\omega\gg\nu_{s,\|}.
     \end{equation}

Similarly, for $\sigma''$ in the case of strongly oblate MNs at low frequencies, one gets
\begin{equation}
 \label{eq 76}
  \left.\sigma''_{\|}=\frac{3}{8\pi}\omega\left(\frac{\omega_{pl}R_{\|}}{\upsilon_F}\right)^2,
   \atop
    \sigma''_{\bot}=\frac{1}{2}\sigma''_{\|}\left(\frac{R_{\bot}}{R_{\|}}
     \arctan{\frac{R_{\bot}}{R_{\|}}}-1\right)\right\},\;\;\;\omega\ll\nu_{s,\bot},
      \end{equation}
and at high frequencies:
\begin{equation}
 \label{eq 77}
  \sigma''_{\bot}\equiv\sigma''_{\|}=\frac{\omega^2_{pl}}{4\pi\omega},\;\;\;\omega\gg\nu_{s,\|}.
   \end{equation}

It remains to consider, how the optical properties of MNs evolve at $\nu_{s,\bot}\leq\omega\leq\nu_{s,\|}$
between the low- and high-frequency interval. This interval is just that for which the parameter
$q_2$ can be greater or less than 1. In this case, only the numerical evaluations of the integrals entered
into Eqs.~(\ref{eq 34}), (\ref{eq 35}) can be performed. The obtained results for real part of $\sigma$
can be found in Ref. [\onlinecite{GT1}].

\subsubsection{Conductivity of a prolate MN}

For prolate particles
\begin{equation}
 \label{eq 78}
  l>2R_{\|}>2R_{\bot}.
   \end{equation}
It is also worth noting that the approximations (\ref{eq 70}), (\ref{eq 71}) still remain to be valid for the case of inequalities (\ref{eq 78}), if the transposition $\nu_{s,\bot}\leftrightarrows\nu_{s,\|}$ is carried out in Eqs.~(\ref{eq 70}), (\ref{eq 71}). As a result, the integrals in Eqs.~(\ref{eq 34}), (\ref{eq 35}) can be easily calculated, and for prolate particles with an arbitrary aspect ratio of $R_{\|}/R_{\bot}$, one obtains the results similar to ones given by Eqs.~(\ref{eq 72}), (\ref{eq 73}), in which the both replacements (\ref{eq 50}) and (\ref{eq 93}) have been made for $J$ and the integrals $I, I^<$. Below, we write down only the results for a strongly prolate particles ($R_{\|}\gg R_{\bot}$) at low and high frequencies. Using Eqs.~(\ref{eq 89}), (\ref{eq 95}), it is easy to find that
\begin{equation}
 \label{eq 79}
  \left.\sigma'_{\|}=\frac{9}{64}\omega^2_{pl}\frac{R_{\bot}}{\upsilon_F},
   \atop
    \sigma'_{\bot}=\sigma'_{\|}/2\right\},\;\;\; \omega\ll\nu_{s,\|},
     \end{equation}
\begin{equation}
 \label{eq 80}
  \left.\sigma'_{\|}=\frac{9}{256}\left(\frac{\omega_{pl}}{\omega}\right)^2\frac{\upsilon_F}{R_{\bot}},
   \atop
    \sigma'_{\bot}=3\sigma'_{\|}/2\right\},\;\;\;\omega\gg\nu_{s,\bot}.
     \end{equation}

For $\sigma''$, just as for $\sigma'$, in the case of strongly prolate MNs, one finds

\begin{equation}
 \label{eq 81}
  \left.\sigma''_{\bot}=\frac{3}{16\pi}\omega\left(\frac{\omega_{pl}R_{\bot}}{\upsilon_F}\right)^2,
   \atop
    \sigma''_{\|}=-2\sigma''_{\bot}\left(1+\ln{\frac{R_{\bot}}{2R_{\|}}}\right)\right\},\;\;\;\omega\ll\nu_{s,\|},
     \end{equation}
and at high frequencies $\omega\gg\nu_{s,\bot}$, we have found $\sigma''_{\|}\simeq\sigma''_{\bot}=\omega^2_{pl}/(4\pi\omega)$,
which coincides exactly with Eq.~(\ref{eq 77}) for oblate particles.

Comparing Eqs.~(\ref{eq 79})--(\ref{eq 81}) for a strongly prolate particle to the corresponding ones (\ref{eq 72})--(\ref{eq 76}) for a strongly oblate particle shows that the character of the frequency dependence of the corresponding components of the electric conductivity tensor remains practically the same. The behavior of this dependence reminds the asymptotic behavior of the Drude-Zommerfeld frequency dependence for an electron scattering in the volume of MN; this can be easily verified from the first summand in the right-hand sides of Eqs. (41), (42) by setting $\omega\ll\nu$ or $\omega\gg\nu$. The difference consists only in the fact that for an electron scattering in the volume, the high-frequency conductivity close to the frequency $\omega\approx\nu$ goes smoothly to the saturation. However, for the strongly asymmetric MNs there exists an entire transitional region between the minimum and the maximum transit rates, where the frequency dependence of $\sigma$ appreciably differs from the volume case.

In the case of the frequency interval $\nu_{s,\|}\leq\omega\leq\nu_{s,\bot}$, the numerical evaluations of the
integrals involved into Eqs.~(\ref{eq 34}), (\ref{eq 35}) can be performed and the obtained results for $\sigma$
have been presented in Ref. [\onlinecite{GT1}].

\section{RESULTS and DISCUSSION}

One of the most important cases takes place for the frequencies $\omega\gg\nu_s$, with $\nu_s=\upsilon_F/(2R)\gg\nu$. In terms of q-language this corresponds to the situation when $|q_1|\ll 1$ and $|q_2|\gg 1$. In this case, one may use the expressions (\ref{eq 58}), (\ref{eq 59}) at $q_1\rightarrow 0$ to obtain the results for an asymptotically high value of $q_2$. As a result, we get for a real and imaginary parts of an electric conductivity of a single spheroidal MN with an arbitrary aspect ratio of $R_{\bot}/R_{\|}$ the following expressions

\begin{equation}
 \label{eq 82}
  \sigma'_{\|\choose\bot}(\omega)\approx\frac{1}{4\pi}\left(\frac{\omega_{pl}}{\omega}\right)^2
   \left(\nu+\frac{9}{4}\frac{\upsilon_F}{R}I_{\|\choose\bot}+\cdots\right),
    \end{equation}
\begin{equation}
 \label{eq 83}
  \sigma''_{\|\choose\bot}(\omega)\approx\frac{1}{4\pi}\left(\frac{\omega_{pl}}{\omega}\right)^2
   \left(\omega-\frac{9}{4}\frac{\upsilon_F}{R}I_{\|\choose\bot}+\cdots\right),
    \end{equation}
provided that $\omega\gg\nu_s\gg\nu$. The parameters $I_{\|}$, $I_{\bot}$ are given in Appendix A by Eqs.~(\ref{eq 86}), (\ref{eq 87}).

For a MN of a spherical shape these expressions take the simplest form
\begin{equation}
 \label{eq 84}
  \sigma'_{q_1\ll 1\choose q_2\gg 1}(\omega)\simeq\frac{3}{8\pi}
   \left(\frac{\omega_{pl}}{\omega}\right)^2\nu_s+\cdots,
    \end{equation}
\begin{equation}
 \label{eq 85}
  \sigma''_{q_1\ll 1\choose q_2\gg 1}(\omega)\simeq\frac{1}{4\pi}
   \left(\frac{\omega_{pl}}{\omega}\right)^2\left(\omega-\frac{3}{2}\nu_s +\cdots\right),
    \end{equation}
provided that $\omega\gg\nu_s\gg\nu$.
The last formula demonstrates simply that the electron interaction with the surface of a spherical
MN can shift the imaginary part of $\sigma$ towards the red side of the frequency scale.
The smaller is the radius of the MN, the greater is the correction to the frequency shift.

Below, we will discuss the size dependence of optical properties of MNs in
more details and will illustrate some results obtained in above sections.

\vskip10pt
\noindent\includegraphics[width=8.6cm]{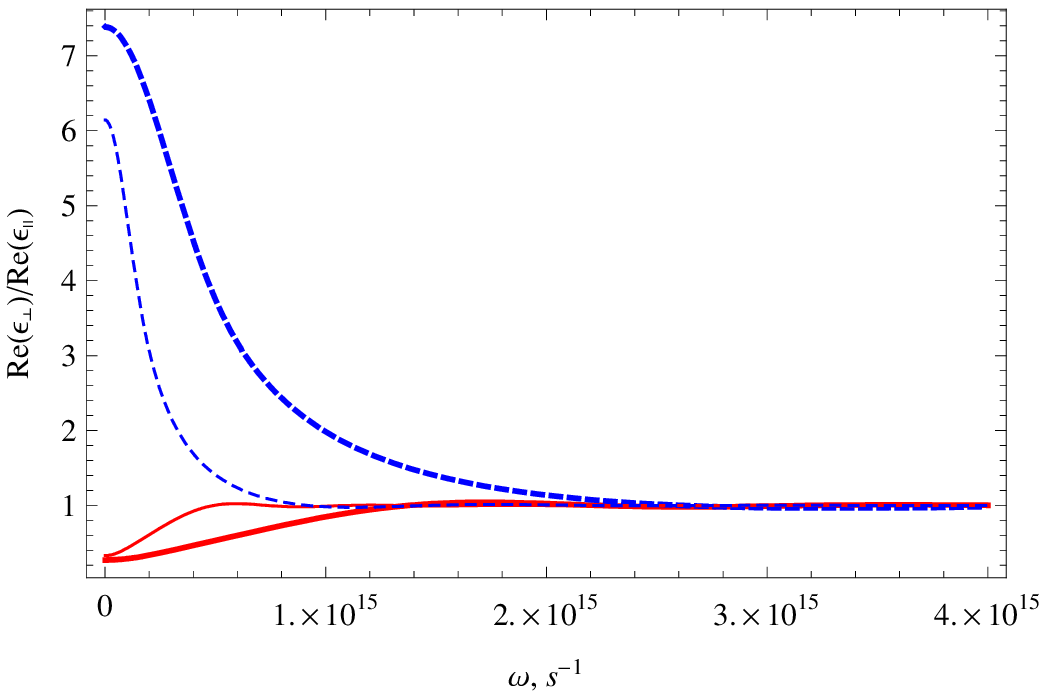}
\vskip-1mm\noindent{\footnotesize FIG.~6. The frequency dependence of the reduced dielectric constant of
Au particles with different $R, \AA$: $50$ (thick lines) and $150$ (thin lines). Full lines are for a prolate
particle with $R_{\bot}/R_{\|}=0.1$, and dashed lines are for an oblate one with $R_{\bot}/R_{\|}=10$.}
\vskip10pt

In Fig.~6, the results of numerical calculations of the ratio between transversal and longitudinal
components of the dielectric permeability versus the frequency are given for Au particles of different radiuses.
The calculations were performed with the use of Eqs.~(\ref{eq 32}), (\ref{eq 35}) and the same numerical parameters as above used.
At high frequencies ($\omega\gg\nu_s$), the ratio of $\epsilon'_{\bot}/\epsilon'_{\|}\rightarrow 1$ for different radiuses of Au particle. This result follows as well from Eqs.~(\ref{eq 32}), (\ref{eq 73}) at $\omega\gg\nu_s$ for particles of an oblate shape,
and from Eqs.~(\ref{eq 32}), (\ref{eq 77}) -- for particles of a prolate shape. At low frequencies ($\omega\ll\nu_s$), the particle
size is strongly effected on the ratio of $Re(\epsilon_{\bot})/Re(\epsilon_{\|})$. As it can be seen in Fig.~6, the curves for MNs of prolate and oblate shapes are merged into one with frequency grows, but for particles of a greater radiuses it becomes at smaller frequencies.

The behavior of $\epsilon'$ as a function of the ellipsoid aspect ratio $R_{\bot}/R_{\|}$ at the frequency of a plasmon
resonance in Au nanoparticle, embedded in the dielectric media with $\epsilon_m=1$, is plotted in Fig.~7. The numerical
calculations with employing of Eqs.~(\ref{eq 32}), (\ref{eq 35}) were performed for two radiuses of a nanoparticle and for two
light polarizations. As one can see in this figure, the longitudinal component of $\epsilon'$ in an oblate MN depends more stronger on the ratio $R_{\bot}/R_{\|}$ than the transverse one. It becomes especially pronounced for Au nanoparticle (with $R=50\AA$) of small oblateness (e.g., $R_{\bot}/R_{\|}<15$), when keeping the laser frequency fixed at $\omega=\omega_{pl}/\sqrt{1+2\epsilon_m}$: the oscillation magnitude of $\epsilon'_{\bot}$ (round the constant of $\epsilon'=-2$) becomes much larger than ones for $\epsilon'_{\|}$.
The amplitude of these oscillations is enhanced and the period is extended with increasing of the ellipsoid aspect ratio $R_{\bot}/R_{\|}$. For MNs of greater radiuses, the number of oscillations is decreased and their period is extended (compare the thin and thick curves in the Fig.~7 for Au particles with $R=50$ and $150\AA$, for example). The both components of $\epsilon'$ no longer oscillate at high oblateness ($R_{\bot}/R_{\|}>80$) of Au particle with $R=50\AA$.

\vskip10pt
\noindent\includegraphics[width=8.6cm]{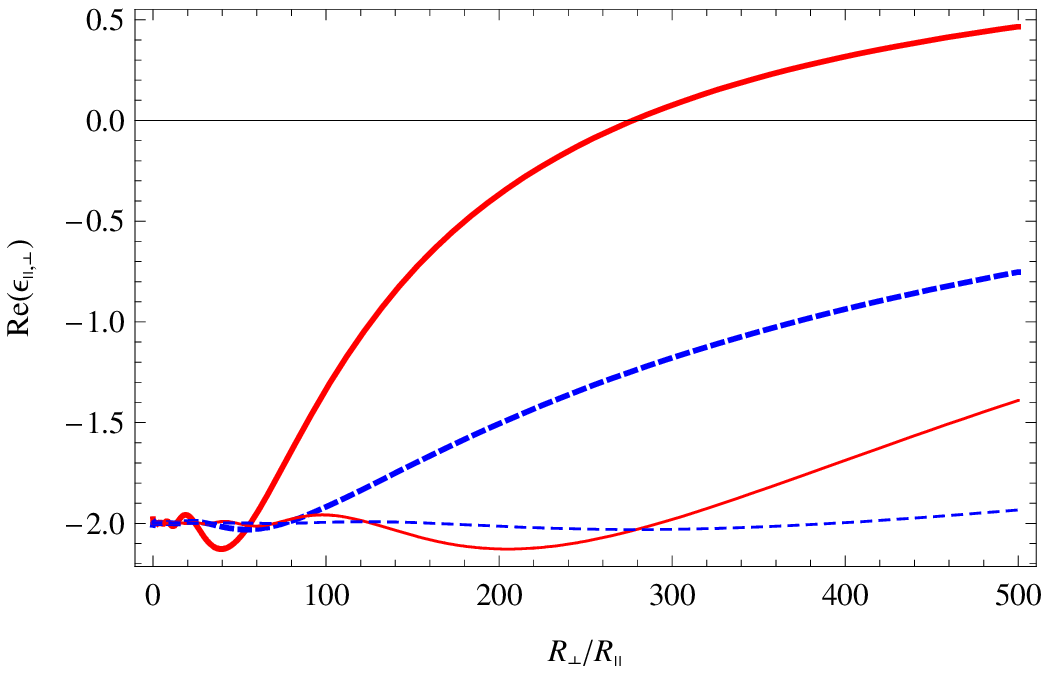}
\vskip-1mm\noindent{\footnotesize FIG.~7. The dependence of the real part of the dielectric constant of Au particles with
different $R, \AA$: $50$ (thick lines) and $150$ (thin lines) on the ellipsoid aspect ratio, at a frequency of plasmon resonance $\omega=\omega_{pl}/\sqrt{3}\simeq 7.91\times 10^{15}\;s^{-1}$.
Solid line is for $\|$- and dashed line for $\bot$-component.}
\vskip10pt

When the size of the particle is large enough, the $\|$- and $\bot$-components of $\epsilon'$ start to come together (see Fig.~2).
For instance, at the frequency $\omega\approx 6\times 10^{14} s^{-1}$ it takes place for Au particle with $R=130\AA$.
At smaller frequencies it occurs for greater radiuses.

Presented dependencies display mainly the behavior of $\epsilon'$ for an oblate nanoparticle ($R_{\bot}/R_{\|}>1$). In the case
of a prolate nanoparticle ($0<R_{\bot}/R_{\|}<1$), the numerical calculations with employing the same Eqs.~(\ref{eq 32}), (\ref{eq 35})
for Au nanoparticle of a small size ($\sim 50\AA$) at the frequency of the plasmon resonance give the oscillations of a transversal
component of $\epsilon'$, the amplitude of which is enhanced and the period is reduced as soon as the prolateness of the MN is increased.
The longitudinal component of $\epsilon'_{\|}$ oscillates in the prolate MN as well, but its amplitude is considerably smaller and the period is much larger than the proper ones for $\epsilon'_{\bot}$.

The imaginary part of the dielectric function as a function of ellipsoid aspect ratio $R_{\bot}/R_{\|}$ is shown in Fig.~8.
The numerical calculations were performed with the use of Eqs.~(\ref{eq 33}), (\ref{eq 34}) at the plasmon frequency for two
radiuses of nanoparticle and for two light polarizations. Since for prolate MN, the value of $\epsilon''$ practically doesn't
depend on the spheroid aspect ratio $R_{\bot}/R_{\|}$ (except for the case of a very high prolateness), in Fig.~8 we present only the
results for an oblate MN. In contrast to the size behavior of $\epsilon'$, described above, one can see that the weak oscillations
of $\epsilon''$ hold together with linear increasing of $\epsilon''$ just at small values of the aspect ratio $R_{\bot}/R_{\|}$. For Au particles with $R=50\AA$ it takes place until $R_{\bot}/R_{\|}<150$, and is more sensible for the longitudinal component of $\epsilon''$ than for the transversal one.
The both components of $\epsilon''$ reach the same maximum at some aspect ratios of $R_{\bot}/R_{\|}$,
which value depends on the radius of particle. In the example depicting in Fig.~8 (for Au particle with $R=50\AA$), the values of $\epsilon''_{\|}=\epsilon''_{\bot}\simeq 1.8$ have peaks at $R_{\bot}/R_{\|}>140$ for $\|$-component and at $R_{\bot}/R_{\|}\simeq 700$
for $\bot$-component of $\epsilon''$. The another interesting feature of the dependence of $\epsilon''$ on $R_{\bot}/R_{\|}$ is the intersection of curves for $\|$- and $\bot$-components for an oblate particle, as it takes place usually for the particle of a spherical shape.

\vskip10pt
\noindent\includegraphics[width=8.6cm]{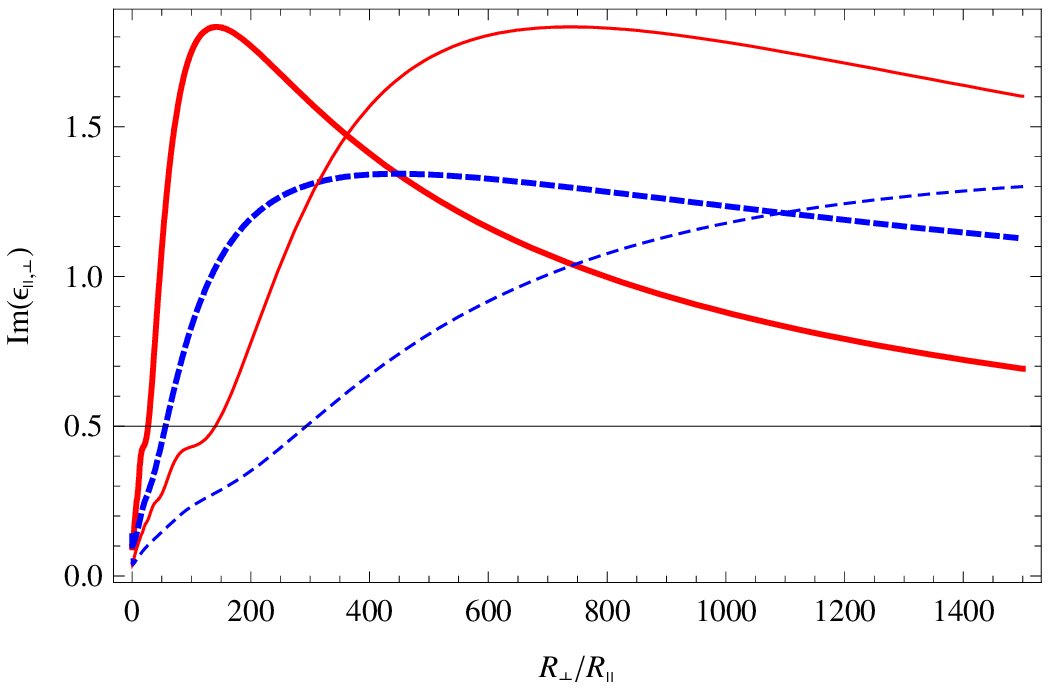}
\vskip-1mm\noindent{\footnotesize FIG.~8. The dependence of the imaginary part of the dielectric constant of Au particles
with different $R, \AA$: $50$ (thick lines) and $150$ (thin lines) on the ellipsoid aspect ratio, at the frequency of a plasmon
resonance $\omega\simeq 7.91\times 10^{15}\;s^{-1}$.
Solid line is for $\|$- and dashed line for $\bot$-component.}
\vskip10pt

Finally, we illustrate, how one may estimate the lifetime of surface plasmon excitation (or any others) in a MN using the
above derived formulas for $\sigma$. As follows from our previous calculations\cite{TG,GT}, the linewidth can be represented as

\begin{equation}
 \label{eq 97}
  \Gamma_{\beta}(\omega)=4\pi L_{\alpha} \sigma_{\alpha\beta}(\omega),
    \end{equation}
where $L_{\alpha}$ is defined above after Eq.~(\ref{eq 6}) and $\sigma_{\alpha\beta}$ is the real part of the conductivity tensor,
given by Eq.~(\ref{eq 23}) for most general situations or by Eq.~(\ref{eq 34}) for the spheroidal MN. For other particular cases, one
may use for $Re\,\sigma$ Eqs.~(\ref{eq 41}), (\ref{eq 47}), (\ref{eq 63}), (\ref{eq 66}), (\ref{eq 72}), (\ref{eq 74}), (\ref{eq 75}), (\ref{eq 79}), (\ref{eq 80}), and (\ref{eq 82}), presented above. In particular, considering only the nanoparticles of a spherical shape ($L=1/3$) and restricting ourselves to the case $\nu\ll\nu_s$, we can choose Eq.~(\ref{eq 63}) for illustration. Substituting it into Eq.~(\ref{eq 97}), we obtain

\begin{equation}
 \label{eq 98}
  \Gamma(\omega)\simeq \frac{\upsilon_F}{4R}
   \left(\frac{\omega_{pl}}{\omega}\right)^2\left[1-
   \frac{2\nu_s}{\omega}\sin\frac{\omega}{\nu_s}+\frac{2\nu^2_s}{\omega^2}\left(1-
    \cos\frac{\omega}{\nu_s}\right)\right],
    \end{equation}
Taking into account only the first term in (\ref{eq 98}), we recover the well-known\cite{KK,YB,MWJ} $1/R$ dependence of $\Gamma$.

\begin{equation}
 \label{eq 99}
  \Gamma_0(\omega, R)=\frac{1}{4}\frac{\upsilon_F}{R}
   \left(\frac{\omega_{pl}}{\omega}\right)^2.
    \end{equation}
As seen from Eqs.~(\ref{eq 98}), (\ref{eq 99}), the lifetime ($1/\Gamma$) of an excitation in the MN depends not only on the nanoparticle radius, but also on the frequency (at which a given excitement is reasonable). For the frequency, which corresponds to the excitation of surface plasmon in MN in a vacuum, $\omega=\omega_{pl}/\sqrt{3}$, the following relation can be obtained from Eq.~(\ref{eq 99}) in energy units:

\begin{equation}
 \label{eq 100}
  \Gamma_0^{SP}(R)=\frac{3}{4}\hbar\frac{\upsilon_F}{R}.
    \end{equation}

The oscillating terms in Eq.~(\ref{eq 98}) give rise to the oscillation of $\Gamma$ around of $\Gamma_0$ as a function
of both the particle radius and the frequency. They can be represented at the frequency of surface plasmon as follows

\begin{eqnarray}
 \label{eq 101}
  &&\Gamma_{osc}^{SP}(R)\simeq \frac{3\sqrt{3}}{4}\frac{\hbar}{\omega_{pl}}\left(\frac{\upsilon_F}{R}
   \right)^2\nonumber \\&\times & \left[-\sin\frac{2R\omega_{pl}}{\sqrt{3}\;\upsilon_F}+
    \frac{\sqrt{3}\;\upsilon_F}{2R\omega_{pl}}\left(1-\cos\frac{2R\omega_{pl}}{\sqrt{3}\;\upsilon_F}\right)\right].
     \end{eqnarray}
The amplitude and period of oscillations can be estimated by the values

\begin{equation}
 \label{eq 102}
  \Gamma^{max}_{osc}\simeq \frac{3\sqrt{3}\,\hbar\upsilon_F^2}{4\omega_{pl}R^2}, \qquad T=\frac{\sqrt{3}\,\pi\upsilon_F}{R\omega_{pl}},
    \end{equation}
accordingly.

It is important to note that in kinetic method this oscillatory behavior of $\Gamma$ follows 
solely from the conditions of an electron scattering on the nanoparticle surface.

Figure 9 shows the full linewidth $\Gamma=\Gamma^{SP}_0+\Gamma^{SP}_{osc}$, which is obtained by numerical evaluating of Eqs.~(\ref{eq 100}), (\ref{eq 101}) for Na nanoparticles with parameters\cite{CK}: $\omega_{pl}=9.18\times 10^{15}$ s$^{-1}$ and $\upsilon_F=1.07\times 10^{8}$ cm/s. One can see that oscillating terms represent an important correction to $\Gamma^{SP}_0(R)$ at small particle radii. This our result for Na nanoparticles only qualitative agrees with the similar results obtained in Refs. [\onlinecite{MWJ},\onlinecite{WMW}]), since we try to apply the kinetic method to the range of $R$, where the quantum effects (like, e.g., the Landau damping) play an important role. But, mostly, for the kinetics, the next inequality could be met

\begin{equation}
 \label{eq 103}
   R\gg\frac{2\pi\hbar}{m\upsilon_F}.
    \end{equation}

In order to study the significance of the oscillatory behavior in more general situations, it is necessary to perform the 
calculations of Eq.~(\ref{eq 97}), using the real parts of Eq.~(\ref{eq 23}) or Eq.~(\ref{eq 34}). This will be done separately.

\vskip10pt
\noindent\includegraphics[width=8.6cm]{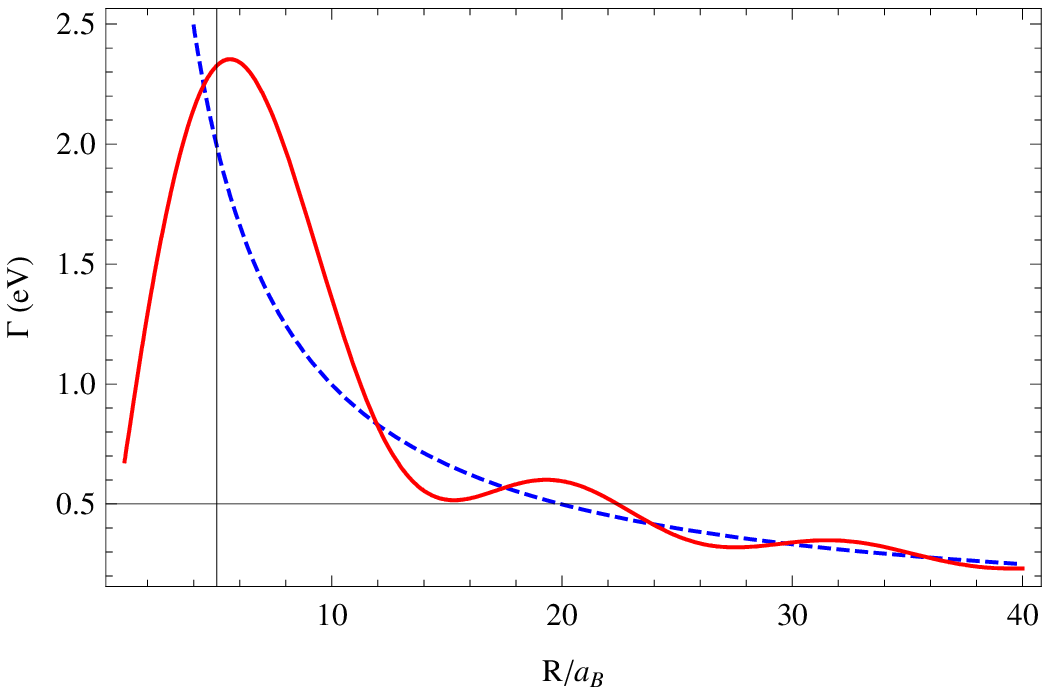}
\vskip-1mm\noindent{\footnotesize FIG.~9. Linewidth $\Gamma(R)$ of the surface plasmon resonance, as a function of radius for Na nanoparticles in units of Bohr radius $a_B\simeq 0.53 \AA$. The smooth term $\Gamma_0(R)$ is given by Eq.~(\ref{eq 100}) (dashed)
and the solid line corresponds to the sum of Eqs.~(\ref{eq 100}) and (\ref{eq 101}).}
\vskip10pt

There are several experimental data for a dielectric constant of powder of Ag and Al,\cite{LCL} and for Ag
nanoparticles.\cite{DCK} For a single Au particle we have found only the experimental data for an optical response.\cite{FCV}
A direct comparison of theoretical results with most of the available experimental measurements of the optical properties of MNs
are still a matter of debate because inhomogeneous in nanoparticle size, shape and local environment hide the homogeneous width of
the surface plasmon resonance. There are very different data even for bulk permittivity of Au,\cite{JC,MJW} especially for its
imaginary part.

\section{CONCLUSIONS}

The kinetic equation method is used to study the peculiarities of the electron interactions with the surface of a spheroidal metal nanoparticle, when the electron scattering from the particle surface becomes a dominant effect. The special attention was paid to study the modification of the Drude-Zommerfeld model applying to the optical properties of MN. The real and imaginary parts of the dielectric permeability at the frequencies above and below the characteristic frequency of a free electron passage between the walls of the particle were calculated for a single oblate and a prolate MN whose dimensions are much smaller than the wavelength of an electromagnetic wave.

It was established for spherical MNs that the kinetic method appreciably changes the frequency dependence of electrical conductivity
at low frequencies, and at high frequencies ($\omega\gg\nu_s$) gives the same result as one obtained from the Drude-Zommerfeld formula.
Quantitatively, the results obtained by kinetic method are of one or even two orders of magnitude lower (depending on the radius of MN) than those following from classical formulas. The difference between results is enhanced markedly as the particle radius is decreased
and the nanoparticle surface starts to play the more pronounced role.

The frequency dependencies of the components of the electric conductivity tensor $\sigma$ were found and their dependence on the
spheroidal aspect ratio was investigated. Simple analytical expressions were found for this tensor in a strongly oblate
or prolate MNs at low and  high frequencies.

The electron interaction with the surface of a spherical
MN can shift the imaginary part of $\sigma$ towards the red side of the frequency scale.
The smaller is the radius of the MN, the greater is the correction to the frequency shift.

Two types of oscillations were established for small enough MNs:
at fixed $\omega$ with varying of $R$, or at fixed $R$ with changing of $\omega$.
These oscillations have a damping character and practically are vanished else at high frequencies or for MNs of high radiuses.

 The ratio of $Im(\epsilon_{\bot})/Im(\epsilon_{\|})$ oscillates with increasing the frequency both for the prolate and the oblate MNs.
 In contrast, the ratio of the real parts of $Re(\epsilon_{\bot})/Re(\epsilon_{\|})$ does not oscillate with frequency. Together with that, the real $\epsilon$ as well as the imaginary $\epsilon$ part of the dielectric permeability oscillate, when the particle radius is increased. It was found that the particle size strongly effects the ratio of $Re(\epsilon_{\bot})/Re(\epsilon_{\|})$ at low frequencies ($\omega\ll\nu_s$).

\begin{acknowledgments}
Authors would like to thank Doctor E.A. Ponezha for her valuable comments and useful remarks.
\end{acknowledgments}

\appendix
\section{}

Below, we present the values for integrals introduced in the Sec. IV.

\begin{eqnarray}
 \label{eq 86}
  I_{\|}&=&\frac{1}{\upsilon}\int\limits_0^{\pi/2}\sin\theta\,\cos^2\theta\,\upsilon'(\theta)\,d\theta =
   \frac{R}{8R_{\|}}\frac{2R_{\bot}^2-R_{\|}^2}{R_{\bot}^2-R_{\|}^2}\nonumber \\&-&
    \frac{R}{8R_{\bot}}\frac{R_{\|}^3}{(R_{\bot}^2-R_{\|}^2)^{3/2}}
     \ln\left|\frac{R_{\bot}}{R_{\|}}+\sqrt{\frac{R_{\bot}^2}{R_{\|}^2}-1}\right|.
      \end{eqnarray}
\begin{eqnarray}
 \label{eq 87}
  I_{\bot}&=&\frac{1}{\upsilon}\int\limits_0^{\pi/2}\frac{1}{2}\sin^3\theta\,\upsilon'(\theta)\,d\theta =
   \frac{3R}{16R_{\|}}-
    \frac{R}{16R_{\|}}\frac{R_{\bot}^2}{R_{\bot}^2-R_{\|}^2}\nonumber \\&+&
     \frac{R}{4R_{\bot}}\frac{R_{\|}}{\sqrt{R_{\bot}^2-R_{\|}^2}}
      \left(1+\frac{R_{\|}^2}{4(R_{\bot}^2-R_{\|}^2)}\right)\nonumber \\&\times&
       \ln\left|\frac{R_{\bot}}{R_{\|}}+\sqrt{\frac{R_{\bot}^2}{R_{\|}^2}-1}\right|.
        \end{eqnarray}

For strongly oblate or prolate MNs, using the Eqs.~(\ref{eq 84}), (\ref{eq 85}), it is easy to find that

\begin{equation}
 \label{eq 88}
  I_{{\|\choose\bot},R_{\bot}\gg R_{\|}}\simeq\frac{1}{4}\frac{R}{R_{\|}}\cdot\left\{1\atop 1/2\right.,
   \end{equation}
\begin{equation}
 \label{eq 89}
  I_{{\|\choose\bot},R_{\bot}\ll R_{\|}}\simeq\frac{\pi}{16}\frac{R}{R_{\bot}}\cdot\left\{1\atop 3/2\right.,
   \end{equation}
correspondingly, and in the limit case of $R_{\bot}=R_{\|}\equiv R$, one gets

\begin{equation}
 \label{eq 90}
  \lim_{R_{\|}\to R_{\bot}} I_{\|}=\lim_{R_{\|}\to R_{\bot}} I_{\bot}=\frac{1}{3}.
   \end{equation}

We advance here as well the another typical integrals, which one meets under calculation of $\sigma$.
\begin{eqnarray}
 \label{eq 91}
  I_{\|}^<&=&\upsilon\int\limits_0^{\pi/2}\frac{\sin\theta\,\cos^2\theta}{\upsilon'(\theta)}\,d\theta =
   \frac{R_{\|}R_{\bot}}{2R(R_{\bot}^2-R_{\|}^2)}\nonumber \\&\times&
    \left[R_{\bot}-\frac{R_{\|}^2}{\sqrt{R_{\bot}^2-R_{\|}^2}}\ln\left|\frac{R_{\bot}}{R_{\|}}+
     \sqrt{\frac{R_{\bot}^2}{R_{\|}^2}-1}\right|\right],
      \end{eqnarray}
\begin{eqnarray}
 \label{eq 92}
  I_{\bot}^<&=&\upsilon\int\limits_0^{\pi/2}\frac{\sin^3\theta}{2\upsilon'(\theta)}\,d\theta =
   -\frac{R_{\|} R_{\bot}^2}{4R (R_{\bot}^2-R_{\|}^2)} \nonumber \\&+&
     \frac{R_{\|}R_{\bot}(2R_{\bot}^2-R_{\|}^2)}{4R(R_{\bot}^2-R_{\|}^2)^{3/2}}
      \ln\left|\frac{R_{\bot}}{R_{\|}}+\sqrt{\frac{R_{\bot}^2}{R_{\|}^2}-1}\right|.
       \end{eqnarray}

In the case of prolate particles ($R_{\|}>R_{\bot}$), one should perform in Eqs.~(\ref{eq 85}), (\ref{eq 86})
and in Eqs.~(\ref{eq 90}), (\ref{eq 91}) the next replacement
\begin{equation}
 \label{eq 93}
  \frac{1}{i}\ln\left|\frac{R_{\bot}}{R_{\|}}+i\sqrt{1-\frac{R_{\bot}^2}{R_{\|}^2}}\right|
   \rightarrow \arcsin{\sqrt{1-\frac{R_{\bot}^2}{R_{\|}^2}}}.
    \end{equation}

For strongly oblate or prolate MNs ($R_{\bot}\gg R_{\|}$), using Eqs.~(\ref{eq 91}), (\ref{eq 92}),
it is easy to obtain that

\begin{equation}
 \label{eq 94}
  I^<_{{\|\choose\bot},R_{\bot}\gg R_{\|}}\simeq\frac{1}{2}\frac{R_{\|}}{R}\cdot\left\{1\atop \ln{2\frac{R_{\bot}}{R_{\|}}-1/2}\right.,
   \end{equation}
\begin{equation}
 \label{eq 95}
  I^<_{{\|\choose\bot},R_{\bot}\ll R_{\|}}\simeq\frac{\pi}{4}\frac{R_{\bot}}{R}\cdot\left\{1\atop 1/2\right.,
   \end{equation}
respectively, and in the limit case of $R_{\bot}=R_{\|}\equiv R$, one finds from Eqs.~(\ref{eq 90}), (\ref{eq 91})

\begin{equation}
 \label{eq 96}
  \lim_{R_{\|}\to R_{\bot}} I_{\|}^<=\lim_{R_{\|}\to R_{\bot}} I_{\bot}^<=\frac{1}{3}.
   \end{equation}

\end{document}